\documentclass[aps,prl,reprint,nobibnotes,nofootinbib]{revtex4-1}
\usepackage[margin=1in]{geometry}
\usepackage{amsmath,amssymb,amsfonts,amsthm,mathtools}
\usepackage{bbm}
\usepackage{mathrsfs}
\usepackage{hyperref}
\usepackage{soul}
\usepackage{slashed}
\usepackage{endnotes}

\renewcommand{\notesname}{Endnotes}

\hypersetup{colorlinks=true,linkcolor=blue,citecolor=blue,urlcolor=blue}
\usepackage{pgfplots}
\pgfplotsset{compat=1.18}
\usetikzlibrary{arrows.meta,positioning,calc}
\newtheorem{theorem}{Theorem}[section]

\newtheorem{corollary}[theorem]{Corollary}

\usepackage{tikz}
\usepackage{tikz-feynman}
\usetikzlibrary{arrows.meta,calc,decorations.pathmorphing,decorations.markings}

\begin{document}

\title{An inquiry into the Absence of Fermion Doubling and why the Nielsen–Ninomiya Theorem Does Not Apply to Nonlocal Quantum Field Theory}

\author{Arvin Kouroshnia}
\affiliation{Department of Physics and Astronomy, University of Waterloo, Waterloo, Ontario N2L 3G1, Canada}

\author{J. W. Moffat}
\affiliation{Perimeter Institute for Theoretical Physics, Waterloo, Ontario N2L 2Y5, Canada}
\affiliation{Department of Physics and Astronomy, University of Waterloo, Waterloo, Ontario N2L 3G1, Canada}

\author{E.~J.~Thompson}
\thanks{Corresponding author (Ethan Thompson)}
\email{thom3471@mylaurier.ca}
\affiliation{Wilfrid Laurier University, Waterloo, Canada, N2L 3C5}

\date{\today}

\begin{abstract}
 In this paper we will examine if nonlocal quantum field theory will suffer from the fermion doubling pathology. We find that for a nonlocal Dirac theory, that no additional fermion species are introduced. This is provided that the form factor is nonvanishing at every point. The proof follows from the invertibility of the entire-function operator, which implies that the nonlocal Dirac operator has exactly the same kernel and finite-momentum zero set as the original local Dirac operator. We will distinguish this result from the standard Nielsen–Ninomiya theorem, which applies local lattice fermions on a compact Brillouin zone. We provide a general criterion for fermion doubling, a test for genuine and false doubling, and then a test procedure for mathematical and physical fermion doubling. We then will go on to distinguish this result from finite derivative truncations, which can introduce spurious polynomial zeros. We then conclude that fermion doubling is absent in the full continuum nonlocal theory.
\end{abstract}

\maketitle

\section{Introductory Considerations}

For decades fermion doubling has been one of the problems that arise when relativistic fermions are formulated on a spacetime lattice when used for quantum theories of gravity. For completeness, we are not saying that fermion doubling only occurs on a spacetime lattice. It also shows up in theories including but not limited to lattice QED, lattice QCD, lattice Electroweak theory and Grand Unified Theories (GUTs)~\cite{Weinberg1967,GeorgiGlashow1974,PatiSalam1974,EichtenPreskill1986,Luscher1999,PoppitzShang2010,GoltermanShamir2024}, lattice Supersymmetry (SUSY)~\cite{DondiNicolai1977,CurciVeneziano1987,Bergner2010,BergnerCatterall2016,Schaich2019}, condensed matter physics~\cite{Wallace1947,Semenoff1984,CastroNeto2009,Armitage2018}, and simulated quantum gravity \& Causal Dynamical Triangulations (CDT)~\cite{LewandowskiZhang2022,BarnettSmolin2015,Barnett2015Thesis,GambiniPullin2015,ZhangLiuHan2022,CatterallLaihoUnmuthYockey2018,Regge1961,AmbjornLoll1998,Loll2020,AmbjornLoll2025}\endnote{If CDT suffers from fermion doubling is still an open question and warrants further research. Some physicists assert that it does suffer the pathology since CDT utilizes discrete simplicial building blocks with a finite cutoff, the underlying momentum space remains bounded. So some suspect that if you freeze a generic triangulation to look at low-energy matter behavior, the spatial discreteness will inevitably trigger unphysical mirror states (taste splitting or doublers). Some do say that CDT does not suffer this issure since in CDT the "lattice" itself is dynamical and highly irregular, and the Nielsen-Ninomiya theorem strictly requires a regular, translation-invariant lattice. Research into random and non-bipartite lattices suggests that structural geometric disorder naturally dampens or "destroys" the clean destructive interference that creates doubler poles, effectively circumventing the theorem. Furthermore, quantum superposition across all possible triangulations may cause unphysical propagation modes to destructively interfere and completely cancel out.}. Fermion doubling appears when a naive finite-difference representation of the Dirac operator does not reproduce a single continuum fermion species, but instead since lattice momentum space is compact and periodic, the lattice kinetic operator develops additional zeros at distinct points of the Brillouin zone. Each of these zeros produces an independent low-energy fermionic excitation, so in four spacetime dimensions a single intended Dirac fermion is replaced by sixteen species. This failure is not just a numerical inconvenience since it alters the particle spectrum (the physical one-particle species or pole content, not the complete mathematical spectrum of the kinetic operator)\endnote{In naive lattice fermion theory, the additional inequivalent zeros of the lattice Dirac operator produce independent low-energy Dirac cones. In four dimensions an intended fermion field then generates sixteen low-energy species, this gets worse in higher dimension theories. This is more than a change in occupation numbers since occupation numbers describe how many particles occupy an already specified set of one-particle states, whereas doubling changes the number of independent fermionic species available to be occupied.}, modifies loop amplitudes and anomaly coefficients, and obstructs a direct lattice realization of chiral gauge theories~\cite{Wilson1974,KogutSusskind1975,DrellWeinsteinYankielowicz1976,Susskind1977,Wilson1977,KarstenSmit1981,MontvayMunster1994,Rothe2012,Smit2002,GattringerLang2010}. The related doubling question has also been discussed in loop-quantum-gravity and quantum-geometry settings by Jacob Barnett, Lee Smolin, and others~\cite{Barnett2015Thesis,BarnettSmolin2015,GambiniPullin2015,LewandowskiZhang2022,ZhangLiuHan2022}.

The mathematical origin of this phenomenon was proved by Nielsen and Ninomiya, who showed that a broad class of local, translation-invariant, Hermitian lattice theories with exact chiral symmetry necessarily contains compensating fermionic modes. Their theorem is topological in character and depends essentially on the compact periodic structure of lattice momentum space together with locality assumptions on the lattice Dirac operator~\cite{NielsenNinomiyaNoGo1981,NielsenNinomiyaI1981,NielsenNinomiyaII1981,Friedan1982,Rebbi1987,ChandrasekharanWiese2004,Kaplan2009}. The standard methods for controlling the problem rely on, including Wilson fermions, staggered fermions, domain-wall fermions, and overlap fermions. Each of these relax or reorganize one or more of the core assumptions of the theorem~\cite{GinspargWilson1982,Kaplan1992,Shamir1993,NarayananNeuberger1994,HasenfratzNiedermayer1994,Neuberger1998a,Neuberger1998b,Luscher1998}. The doubling problem must be understood as a statement about the zero structure of a particular class of lattice kinetic operators, and not just as a universal property of every ultraviolet modification of fermionic quantum field theory~\cite{KarstenSmit1978,KarstenSmit1979,BodwinKovacs1987,CampostriniCurciPelissetto1987,Pelissetto1988}.

In the class of theories we considered here, the ultraviolet behavior is modified not by replacing the continuum spacetime with a discrete lattice, but instead by acting on the fields or kinetic operators with an entire function of a covariant differential operator, with this the resulting theory remains formulated on a continuum manifold, but the nonlocal form factor suppresses large Euclidean momenta and introduces a fundamental resolution scale. If the entire function is nonvanishing at every finite argument (point $x$ in the function $f(x)$), then it defines an invertible operator on the relevant field space. These deformations\footnote{Multiplying the Dirac operator by the nonlocal entire-function factor changes how strongly different momentum modes contribute, especially at high momentum.} (a systematic modification of the original kinetic operator) will change the ultraviolet weight of fermionic modes but it does not obviously alter the zero set of the underlying Dirac operator~\cite{Efimov1967,AlebastrovEfimov1973,Krasnikov1987,Moffat1990,EvensMoffatKleppeWoodard1991,KleppeWoodard1992,Tomboulis1997,BarnabyKamran2008,BiswasGerwickKoivistoMazumdar2012,ModestoRachwal2017,BuoninfanteLambiaseMazumdar2019}.

Despite how the nonlocal regulator\footnote{A regulator is a mathematical modification introduced to control ultraviolet divergences or otherwise make ill-defined expressions finite. We should note there are two different kinds, for local QFT and in conventional regularization, the regulator is just a computational device. One introduces a cutoff $\Lambda$, renormalizes the theory, and then attempts to remove the regulator ($\Lambda\to\infty$). In the Moffat–Thompson framework, $E_M$ is a physical nonlocality scale, so the entire-function factor is not merely a temporary regulator. It is part of the fundamental kinetic operator and changes the theory’s ultraviolet behavior.} is implemented. it has previously been asserted that the earlier claim that nonlocal regularization avoids fermion doubling was not correct. In a comment by James Cline~\cite{Cline2025,ClineComment,MoffatThompsonReply2025}, there was a brief footnote stating that the claimed circumvention of the doubling problem first presented in 1991 had failed, citing a private communication from Richard P. Woodard. After an email exchange with James Cline, he noted that he misinterpreted Richard P. Woodards comment and therefore what he said was not accurate to what Woodard said. Despite wrongly writing a comment based on something that ended up being a big misinterpretation there was no explicit nonlocal Dirac operator, additional fermionic zero, propagator pole, or Brillouin-zone analysis provided in support of that assertion. So we believe that this statement raises an interesting mathematical question that should be settled directly at the level of the fermion kinetic operator.

The proof presented later in the paper is trivial in nature but is important to settle this issue and question, we show that fermion doubling is identified with a multiplicity of inequivalent zeros of the inverse fermion propagator. An invertible entire-function factor cannot create new zeros for the operator it multiplies, so the kernel of the nonlocal Dirac operator is identical to the kernel of the ordinary continuum Dirac operator, and the finite-momentum zero set is unchanged. In the translation-invariant vacuum the determinant of the nonlocal kinetic operator vanishes only on the usual Dirac mass shell, while the corresponding propagator contains no additional one-particle poles.

A final distinction we present concerns the expansion of the entire function in powers of the covariant d'Alembertian, where the full infinite series defines a single analytic operator and must not be confused with an arbitrary finite polynomial truncation. A finite truncation may possess zeros that are absent from the exact entire function and can therefore generate false higher-derivative modes that are not part of the theory. Such modes are artifacts of replacing the original nonlocal theory by a different finite-order theory and they are not fermion doublers in the Nielsen–-Ninomiya sense, and they do not invalidate the spectral result for the complete zero-free entire operator. There are related applications of nonlocal quantum-field-theoretic methods to covariance, causality, localization, Wick ordering, phenomenology, gravity, string-field observables, and Ostrogradsky-type questions are discussed in~\cite{MoffatThompsonRegulators2026,ThompsonCovariance2026,ThompsonMicrocausality2026,ThompsonAsymptoticLocality2026,ThompsonMacrocausality2026,ThompsonLocalization2026,ThompsonWickOrdering2026,ThompsonKouroshniaSFT2026,ThompsonTopQuark2025,ThompsonDeSitterCores2026,ThompsonGWEchoes2026,ThompsonGluonFusion2026,ThompsonOstrogradsky2026}.

\section{The Ordinary Doubling Problem and the Nielsen–-Ninomiya Theorem}

We first will review where fermion doubling in a lattice theory comes from, so that the proof that nonlocal quantum field theory does not suffer this issue is clear. Since the issue is most clearly understood as a statement about the zeros and topology of the lattice Dirac operator, the doubling phenomenon is topologically protected in local theories, but the breaking of strict locality allows the Dirac operator's eigenvalues to evade these topological constraints. We will work in \(d\) Euclidean spacetime dimensions, with lattice spacing taken to be a positive value \(a>0\), lattice sites given by \(x\in a\mathbb{Z}^{d}\), and unit lattice vectors \(\hat\mu\), where \(\mu=1,\ldots,d\) labels the spacetime directions.

For a fermion field \(\psi(x)\) the naive symmetric lattice derivative is defined by:
\begin{equation}
\nabla^{(a)}_{\mu}\psi(x)
=
\frac{\psi(x+a\hat\mu)-\psi(x-a\hat\mu)}{2a},
\label{eq:symmetric-lattice-derivative}
\end{equation}
where \(a\) is the distance between neighboring lattice points, and \(\nabla^{(a)}_{\mu}\) approximates the continuum derivative \(\partial_\mu\) in the limit \(a\rightarrow 0\). The corresponding massless naive lattice Dirac operator is:
\begin{equation}
D_{\rm lat}
=
\sum_{\mu=1}^{d}\gamma_\mu\nabla^{(a)}_{\mu},
\label{eq:naive-lattice-Dirac}
\end{equation}
where the matrices \(\gamma_\mu\) are Euclidean Dirac gamma matrices, they are $4\times4$ complex matrices that act on the spin components of fermion fields, the gamma matrices satisfy the Euclidean Clifford algebra:
\begin{equation}
\{\gamma_\mu,\gamma_\nu\}
=
2\delta_{\mu\nu}\mathbf{1},
\label{eq:Euclidean-clifford}
\end{equation}
where in this equation, \(\delta_{\mu\nu}\) is the Kronecker delta and \(\mathbf{1}\) is the identity matrix on spinor space.

We now act on a plane-wave mode:
\begin{equation}
\psi(x)=u(p)e^{ip\cdot x},
\label{eq:lattice-plane-wave}
\end{equation}
where \(p=(p_1,\ldots,p_d)\) is the lattice momentum, \(u(p)\) is a spinor amplitude, and \(p\cdot x=\sum_{\mu}p_\mu x_\mu\). Plugging Eq.~\eqref{eq:lattice-plane-wave} into Eq.~\eqref{eq:symmetric-lattice-derivative} gives us:
\begin{equation}
\nabla^{(a)}_{\mu}\psi(x)
=
i\frac{\sin(ap_\mu)}{a}\psi(x).
\label{eq:lattice-derivative-momentum}
\end{equation}
So the momentum-space Dirac operator is given by:
\begin{equation}
D_{\rm lat}(p)
=
\frac{i}{a}
\sum_{\mu=1}^{d}\gamma_\mu\sin(ap_\mu),
\label{eq:lattice-Dirac-momentum}
\end{equation}
so equation~\eqref{eq:lattice-Dirac-momentum} shows us that the continuum momentum component \(p_\mu\) has been replaced by the periodic function \(\sin(ap_\mu)/a\).

Since the lattice is discrete in position space, when we are in momentum space it is compact and periodic. We note that it is sufficient to restrict each momentum component to the first Brillouin zone:
\begin{equation}
-\frac{\pi}{a}
<
p_\mu
\leq
\frac{\pi}{a},
\label{eq:Brillouin-zone}
\end{equation}
where the full momentum space is therefore the \(d\)-dimensional torus $\mathcal{B}\simeq T^{d}$. The identification as a torus means that opposite faces of the Brillouin zone are physically equivalent because momenta differing by \(2\pi/a\) will represent the same lattice mode.

A massless fermionic excitation occurs at every zero of \(D_{\rm lat}(p)\), and from Eq.~\eqref{eq:lattice-Dirac-momentum}, this requires:
\begin{equation}
\sin(ap_\mu)=0,
\qquad
\forall\mu.
\label{eq:zero-condition}
\end{equation}
Within the Brillouin zone we note that each component satisfies the conditions that either:
\begin{equation}
p_\mu
=
0,
\qquad
\text{or}
\qquad
p_\mu
=
\frac{\pi}{a}.
\label{eq:component-zeros}
\end{equation}
There are two choices for each of the \(d\) momentum components, so the total number of zeros is:
\begin{equation}
N_{\rm zero}=2^{d},
\label{eq:number-of-doublers}
\end{equation}
in four spacetime dimensions this gives us:
\begin{equation}
N_{\rm zero}=2^{4}=16,
\label{eq:sixteen-species}
\end{equation}
so we see that the naive lattice theory therefore describes sixteen low-energy fermionic modes rather than one.

We can see that each zero behaves as an independent continuum fermion if we label the zeros by:
\begin{equation}
n_\mu\in\{0,1\},
\qquad
p^{(n)}_\mu=\frac{\pi n_\mu}{a},
\label{eq:corner-label}
\end{equation}
and expand around one of the zeros:
\begin{equation}
p_\mu=p^{(n)}_\mu+k_\mu,
\qquad
|ak_\mu|\ll 1,
\label{eq:corner-expansion}
\end{equation}
where \(k_\mu\) is the small momentum measured relative to the chosen zero, using:
\begin{align}
\notag\sin(\pi n_\mu+ak_\mu)
=
(-1)^{n_\mu}\sin(ak_\mu)
\\=
(-1)^{n_\mu}ak_\mu
+
O(a^{3}k_\mu^{3}),
\label{eq:sine-expansion}
\end{align}
we can obtain:
\begin{equation}
D_{\rm lat}(p)
=
i\sum_{\mu=1}^{d}
(-1)^{n_\mu}\gamma_\mu k_\mu
+
O(a^{2}k^{3}).
\label{eq:Dirac-near-corner}
\end{equation}
If we define:
\begin{equation}
\gamma^{(n)}_\mu
=
(-1)^{n_\mu}\gamma_\mu,
\label{eq:effective-gamma}
\end{equation}
then the matrices \(\gamma^{(n)}_\mu\) satisfy the same Clifford algebra as the original \(\gamma_\mu\), and this means that Eq.~\eqref{eq:Dirac-near-corner} is again a continuum Dirac operator to leading order. Every zero of the lattice operator therefore generates an independent light fermion species~\cite{Susskind1977,KarstenSmit1981,MontvayMunster1994,Rothe2012,Smit2002}.

The topological content of this result is given by the Nielsen–-Ninomiya theorem, in its standard form, the theorem says that a lattice Dirac operator cannot describe a single isolated chiral fermion if it satisfies all of the following conditions: the first being translation invariance, the second being locality, the third being Hermiticity, the fourth being exact chiral symmetry, and the fifth being the correct continuum behavior near its zeros. The theorem assumes that momentum space is the compact torus \(T^{d}\), and it constrains the topology of the map defined by the lattice Dirac operator over that torus~\cite{NielsenNinomiyaNoGo1981,NielsenNinomiyaI1981,NielsenNinomiyaII1981,Friedan1982}.

For a massless lattice fermion we write the Dirac operator in its general form:
\begin{equation}
D(p)
=
i\sum_{\mu=1}^{d}\gamma_\mu f_\mu(p),
\label{eq:general-lattice-Dirac}
\end{equation}
where each \(f_\mu(p)\) is a real and periodic function of momentum. Away from the zeros of \(D(p)\) we can define the normalized vector field:
\begin{equation}
\widehat f_\mu(p)
=
\frac{f_\mu(p)}
{\sqrt{\sum_{\nu=1}^{d}f_\nu(p)^2}},
\label{eq:normalized-vector-field}
\end{equation}
where the vector \(\widehat f(p)\) lies on the unit sphere \(S^{d-1}\). So the lattice Dirac operator defines away from its zeros, a map from the momentum torus into the sphere of normalized Dirac coefficients.

Each isolated zero \(p^{(i)}\) carries an integer topological index:
\begin{equation}
\nu_i
=
\operatorname{sgn}
\det
\left[
\frac{\partial f_\mu}{\partial p_\nu}
\right]_{p=p^{(i)}},
\label{eq:topological-index}
\end{equation}
so long as the Jacobian matrix is nonsingular at the zero. The matrix in Eq.~\eqref{eq:topological-index} measures how the vector \(f_\mu(p)\) changes in the neighborhood of the zero, while the sign of its determinant determines the orientation of the corresponding Dirac cone, and in a chiral theory this sign is related to the effective chirality carried by that zero.

Because the Brillouin zone is compact and has no boundary the total topological index must vanish:
\begin{equation}
\sum_i \nu_i=0,
\label{eq:index-sum}
\end{equation}
this equation is the essential topological statement behind fermion doubling, basically it means if a single zero with nonzero index cannot exist by itself then it must be accompanied by one or more additional zeros whose indices cancel it. So left and right handed lattice fermionic modes necessarily occur with equal total topological charge.

The theorem then can be summarized as follows~\cite{NielsenNinomiyaNoGo1981,NielsenNinomiyaI1981,NielsenNinomiyaII1981,ChandrasekharanWiese2004,Kaplan2009}.

\begin{theorem}[Nielsen–-Ninomiya theorem.]
A local, translation-invariant, Hermitian lattice Dirac operator with exact chiral symmetry and the correct continuum limit cannot contain a single isolated chiral fermion. Its zeros occur in topologically compensating sets whose total index vanishes.
\end{theorem}

\medskip

Now we know that this theorem does not state that every nonlocal fermion theory must double, as it applies to a specific class of lattice operators defined on compact periodic momentum space. In particular we note that the two assumptions most relevant for the present work are the compact Brillouin torus and locality of the lattice Dirac operator. The class of nonlocal quantum field theory we study has neither a compact momentum torus nor a local position-space kinetic operator, so it obviously lies outside the hypotheses of the theorem.

This observation alone does not yet prove the absence of doubling in nonlocal quantum field theory as it only shows that the Nielsen–-Ninomiya theorem does not force it. The actual proof must be obtained directly by comparing the zeros and kernels of the local and nonlocal Dirac operators.

\section{The Continuum Nonlocal Dirac Operator}

We now will define the class of nonlocal fermion theories for which the absence of doubling will be proved, we want to note that the key distinction from the lattice construction is that spacetime and momentum space remain continuous. Nonlocality is introduced through an entire function of a covariant differential operator rather than through a finite-difference replacement of the derivative~\cite{Efimov1967,AlebastrovEfimov1973,Krasnikov1987,Moffat1990,EvensMoffatKleppeWoodard1991,KleppeWoodard1992}.

We let \(\psi(x)\) be a Dirac spinor field on four-dimensional Minkowski spacetime and let:
\begin{equation}
\bar\psi(x)
=
\psi^\dagger(x)\gamma^0
\label{eq:Dirac-adjoint}
\end{equation}
be its Dirac adjoint. Here \(\psi^\dagger\) is the Hermitian conjugate of the spinor and \(\gamma^0\) is the timelike Dirac matrix. The gamma matrices satisfy the Minkowski-Clifford algebra:
\begin{equation}
\{\gamma^\mu,\gamma^\nu\}
=
2\eta^{\mu\nu}\mathbf{1},
\label{eq:Minkowski-clifford}
\end{equation}
here \(\eta^{\mu\nu}\) is the inverse Minkowski metric, \(\mathbf{1}\) is the identity matrix on spinor space as before, and Greek indices \(\mu,\nu=0,1,2,3\) label spacetime coordinates.

For a fermion transforming in a representation of a gauge group, the gauge-covariant derivative is:
\begin{equation}
D_\mu
=
\partial_\mu+A_\mu,
\label{eq:gauge-covariant-derivative}
\end{equation}
where \(\partial_\mu\) is the ordinary spacetime derivative and \(A_\mu\) is the gauge connection acting in the fermion representation, and coupling constants and Lie-algebra generators can be absorbed into \(A_\mu\). The corresponding covariant Dirac operator is:
\begin{equation}
\mathcal{D}_0
=
i\gamma^\mu D_\mu-m,
\label{eq:local-Dirac-operator}
\end{equation}
where \(m\geq 0\) is the fermion mass. The equation:
\begin{equation}
\mathcal{D}_0\psi=0
\label{eq:local-Dirac-equation}
\end{equation}
is the ordinary gauge-covariant Dirac equation where $\mathcal{D}_0$ encodes the energy or time evolution or static state constraint of the system.

To construct our nonlocal theory we define the covariant second-order operator:
\begin{equation}
\Box_D
=
D^\mu D_\mu,
\label{eq:covariant-box}
\end{equation}
this operator \(\Box_D\) is the gauge-covariant d'Alembertian acting on spinor fields. Its precise sign depends on the metric convention but for the proof below it is independent of that convention so long as it is used consistently. We let:
\begin{equation}
F:\mathbb{C}\longrightarrow\mathbb{C},
\label{eq:entire-map}
\end{equation}
we can also write this map as an endomorphism as it is a map that takes a mathematical object back to itself:
\begin{equation}
F \in \text{End}_{\text{hol}}(\mathbb{C}),
\end{equation}
be an entire function, here entire means that \(F(z)\) is complex differentiable at every finite \(z\in\mathbb{C}\), and therefore possesses a globally convergent power-series expansion:
\begin{equation}
F(z)
=
\sum_{n=0}^{\infty}c_n z^n,
\label{eq:entire-series}
\end{equation}
where the constants \(c_n\) are the Taylor coefficients of \(F\). We can then impose the admissibility conditions:
\begin{equation}
F(0)=1,
\qquad
F(z)\neq 0
\quad
\forall\;\;\text{finite }z\in\mathbb{C}.
\label{eq:admissible-form-factor}
\end{equation}
the condition of \(F(0)=1\) will ensure recovery of the ordinary local theory at momenta small compared with the nonlocal scale. The zero-free condition ensures that multiplication by \(F\) is invertible and that the form factor does not introduce additional finite-momentum singularities.

The nonlocality scale is represented by a positive energy scale called the Moffat energy $E_M>0$, where \(E_M\) has dimensions of energy or mass. The associated characteristic length is of order \(E_M^{-1}\) in units with \(\hbar=c=1\). The dimensionless operator entering the form factor is thus \(\Box_D/E_M^2\).

Using analytic functional calculus the operator-valued form factor is defined by~\cite{ReedSimonI1980}:
\begin{equation}
F\left(\frac{\Box_D}{E_M^2}\right)
=
\sum_{n=0}^{\infty}
c_n
\left(\frac{\Box_D}{E_M^2}\right)^n,
\label{eq:operator-entire-series}
\end{equation}
this equation contains arbitrarily high powers of the covariant d'Alembertian and is thus nonlocal in position space. It must be understood as the complete infinite series, not as an arbitrarily high finite polynomial truncation or that will cause problems that we will go over later~\cite{BarnabyKamran2008,BuoninfanteLambiaseMazumdar2019}.

We define the nonlocal Dirac operator by:
\begin{equation}
\mathcal{D}_F
=
F^{-1}\left(\frac{\Box_D}{E_M^2}\right)\mathcal{D}_0,
\label{eq:nonlocal-Dirac-operator}
\end{equation}
where \(F^{-1}\) is the reciprocal function which is entire because \(F(z)\) has no finite zeros. The factor in Eq.~\eqref{eq:nonlocal-Dirac-operator} acts on the result of \(\mathcal{D}_0\), this ordering is stated explicitly because in a nontrivial gauge background \(\Box_D\) does not need to commute with \(\mathcal{D}_0\). The kernel proof requires only that the operator:
\begin{equation}
A_F
=
F^{-1}\left(\frac{\Box_D}{E_M^2}\right),
\label{eq:invertible-factor}
\end{equation}
be invertible on the common domain of the operators. The nonlocal equation of motion is given by:
\begin{equation}
\mathcal{D}_F\psi=0,
\label{eq:nonlocal-Dirac-equation}
\end{equation}
and because \(\mathcal{D}_F=A_F\mathcal{D}_0\), the nonlocal modification multiplies the local Dirac equation by an invertible operator, this structure will be the basis of the no-doubling theorem.

For the explicit momentum-space analysis we will specialize to the translation-invariant vacuum where:
\begin{align}
A_\mu&=0,
\\
D_\mu&=\partial_\mu,
\\
\Box_D&=\Box.
\label{eq:flat-vacuum}
\end{align}
A plane-wave spinor is written as:
\begin{equation}
\psi(x)
=
u(p)e^{-ip\cdot x},
\label{eq:continuum-plane-wave}
\end{equation}
where \(p_\mu\in\mathbb{R}^{1,3}\) is the continuous four-momentum and \(u(p)\) is a four-component spinor amplitude. Acting on this mode gives us:
\begin{align}
\Box\psi(x)
&=
-p^2\psi(x),
\\
p^2&=\eta^{\mu\nu}p_\mu p_\nu.
\label{eq:box-eigenvalue}
\end{align}
So we see that the entire operator becomes an ordinary scalar multiplier:
\begin{equation}
F\left(\frac{\Box}{E_M^2}\right)\psi(x)
=
F\left(-\frac{p^2}{E_M^2}\right)\psi(x).
\label{eq:momentum-form-factor}
\end{equation}
Then the momentum-space nonlocal Dirac operator is:
\begin{equation}
\mathcal{D}_F(p)
=
F^{-1}\left(-\frac{p^2}{E_M^2}\right)
(\slashed p-m),
\label{eq:momentum-nonlocal-Dirac}
\end{equation}
where:
\begin{equation}
\slashed p
=
\gamma^\mu p_\mu
\label{eq:slash-momentum}
\end{equation}
is just the usual Dirac slash notation.

Equation~\eqref{eq:momentum-nonlocal-Dirac} shows us the crucial difference from the naive lattice operator as the continuum momentum \(p_\mu\) has not been replaced by a periodic trigonometric function, but instead the ordinary Dirac operator is multiplied by a scalar function of the Lorentz invariant \(p^2\). Since the multiplier is nonzero for every finite momentum, it cannot by itself create additional zeros of the Dirac operator.

\section{A General Theorem on Invertible Deformations of Kinetic Operators}

The argument we just used above is not just for nonlocal quantum field theory and it is not actually specific to fermions, it just follows from the general operator-theoretic fact that multiplication of a kinetic operator by an injective operator cannot create or remove solutions of the corresponding homogeneous field equation. We can formulate this statement carefully before returning to its effect for particle spectra, that being fermions.

We let \(X\) and \(Y\) be complex vector spaces, which in field theory may be taken to be suitable spaces of classical fields or distributions, then we let:
\begin{equation}
K:\mathcal{D}(K)\subseteq X\longrightarrow Y
\label{eq:general-K}
\end{equation}
be a linear operator with domain \(\mathcal{D}(K)\). The operator \(K\) could represent a kinetic operator, wave operator, linearized field equation, constraint operator, or inverse free propagator depending on what one is working on. Its kernel is defined by:
\begin{equation}
\ker K
=
\left\{
\phi\in\mathcal{D}(K)\;:\;K\phi=0
\right\}.
\label{eq:general-kernel}
\end{equation}
So \(\ker K\) is the set of all fields \(\phi\) that solve the homogeneous equation \(K\phi=0\).

We now introduce a second linear operator:
\begin{equation}
A:\mathcal{D}(A)\subseteq Y\longrightarrow Y,
\label{eq:general-A}
\end{equation}
and assume that the range of \(K\):
\begin{equation}
\operatorname{Ran}K
=
\left\{
K\phi\;:\;\phi\in\mathcal{D}(K)
\right\},
\label{eq:general-range}
\end{equation}
is contained in the domain of \(A\), this condition will ensure that the composition \(AK\) is well defined. We then define the deformed operator:
\begin{equation}
K_A=AK,
\label{eq:general-deformation}
\end{equation}
this means that \(K\) first acts on the field \(\phi\), after which \(A\) acts on the result \(K\phi\).

The minimal condition required for preservation of the kernel is that \(A\) be injective on the range of the operator (or matrix) \(K\), written as: \(\operatorname{Ran}K\). This means that:
\begin{equation}
y\in\operatorname{Ran}K,
\qquad
Ay=0
\quad\Longrightarrow\quad
y=0.
\label{eq:injective-range}
\end{equation}
Full invertibility of \(A\) is sufficient for Eq.~\eqref{eq:injective-range}, but it is stronger than necessary.

\begin{theorem}[General Theorem on Invertible Deformations of Kinetic Operators]
Let \(K:\mathcal{D}(K)\subseteq X\rightarrow Y\) be a linear operator and let \(A\) be a linear operator whose domain contains \(\operatorname{Ran}K\). If \(A\) is injective on \(\operatorname{Ran}K\), then
\begin{equation}
\ker(AK)=\ker K.
\label{eq:kernel-invariance-general}
\end{equation}
\end{theorem}

\medskip
\noindent We can first prove the inclusion:
\begin{equation}
\ker K\subseteq\ker(AK).
\label{eq:first-inclusion-general}
\end{equation}
First let \(\phi\in\ker K\), then by the definition of the kernel:
\begin{equation}
K\phi=0,
\label{eq:Kphi-zero-general}
\end{equation}
by acting with \(A\) on both sides gives us:
\begin{equation}
AK\phi=A0=0,
\label{eq:AKphi-zero-general}
\end{equation}
where linearity implies \(A0=0\). Therefore \(\phi\in\ker(AK)\), proving Eq.~\eqref{eq:first-inclusion-general}.

We next can prove the reverse inclusion:
\begin{equation}
\ker(AK)\subseteq\ker K,
\label{eq:second-inclusion-general}
\end{equation}
now let \(\phi\in\ker(AK)\), then we see:
\begin{equation}
AK\phi=0.
\label{eq:AK-zero-converse}
\end{equation}
And since \(K\phi\in\operatorname{Ran}K\), the injectivity condition \eqref{eq:injective-range} implies:
\begin{equation}
K\phi=0.
\label{eq:K-zero-converse}
\end{equation}
So we have \(\phi\in\ker K\), which proves Eq.~\eqref{eq:second-inclusion-general}. Combining the two inclusions establishes Eq.~\eqref{eq:kernel-invariance-general}.

\medskip

This theorem does not require that \(A\) commute with \(K\), just note that the ordering in Eq.~\eqref{eq:general-deformation} must only be specified consistently, as in particular if \(A\) is invertible on \(Y\), then Eq.~\eqref{eq:kernel-invariance-general} follows immediately from:
\begin{equation}
AK\phi=0
\quad\Longrightarrow\quad
A^{-1}AK\phi=0
\quad\Longrightarrow\quad
K\phi=0,
\label{eq:inverse-proof-general}
\end{equation}
where \(A^{-1}\) is the inverse operator satisfying \(A^{-1}A=1_Y\), where \(1_Y\) is the identity operator on \(Y\).

The theorem shows that an injective left multiplication cannot introduce new zero modes, but it does not by itself imply that every nonzero eigenvalue of \(K\) is unchanged. The kernel preservation is weaker than complete spectral equivalence.

For a translation-invariant theory the kinetic operator may be represented in momentum space by a finite-dimensional matrix:
\begin{equation}
K(p):V\longrightarrow V,
\label{eq:momentum-K}
\end{equation}
where \(p\) is the continuum momentum and \(V\) is the finite-dimensional internal vector space carrying either spinor, Lorentz, flavour, or gauge indices. For example for a four-component Dirac field we have \(V\simeq\mathbb{C}^{4}\)~\cite{PeskinSchroeder1995,WeinbergQFTI1995}.

We let \(A(p):V\rightarrow V\) be a matrix-valued function of momentum, and we define:
\begin{equation}
K_A(p)=A(p)K(p).
\label{eq:momentum-deformation}
\end{equation}
Taking the determinant gives us:
\begin{equation}
\det K_A(p)
=
\det A(p)\,\det K(p),
\label{eq:det-product-general}
\end{equation}
this follows directly from the multiplicativity of the determinant for finite-dimensional matrices.

Now suppose that:
\begin{equation}
\det A(p)\neq 0
\label{eq:A-nonsingular}
\end{equation}
for every finite momentum \(p\) in the momentum domain under consideration, then we have that \(A(p)\) is invertible at every such momentum, and Eq.~\eqref{eq:det-product-general} implies:
\begin{equation}
\det K_A(p)=0
\quad\Longleftrightarrow\quad
\det K(p)=0,
\label{eq:zero-set-invariance}
\end{equation}
so this tells us that the deformed and undeformed kinetic operators possess the same finite-momentum zero set.

If the free propagator is defined by:
\begin{equation}
G(p)=iK^{-1}(p),
\label{eq:general-propagator}
\end{equation}
at momenta where \(K(p)\) is invertible, then the deformed propagator is given by:
\begin{equation}
G_A(p)
=
iK_A^{-1}(p)
=
iK^{-1}(p)A^{-1}(p).
\label{eq:general-deformed-propagator}
\end{equation}
The ordering in Eq.~\eqref{eq:general-deformed-propagator} follows from:
\begin{equation}
\bigl(AK\bigr)^{-1}=K^{-1}A^{-1},
\label{eq:inverse-product-order}
\end{equation}
so if both \(A(p)\) and \(A^{-1}(p)\) are finite and analytic at every finite momentum, then they cannot introduce additional finite poles into the regulated propagator \(G_A(p)\). The locations of the free one-particle poles are therefore determined by the same zeros of \(\det K(p)\) as in the undeformed theory.

The multiplier \(A(p)\) may still modify the residues of the propagator, its ultraviolet behavior, and its position-space support. But the theorem preserves the finite zero and pole sets, but it does not state that the two theories have identical off-shell Green functions.

A more general deformation may act on both sides of the kinetic operator:
\begin{equation}
K_{A,B}=AKB,
\label{eq:two-sided-deformation}
\end{equation}
where \(A\) and \(B\) are invertible operators with compatible domains. The equation:
\begin{equation}
AKB\phi=0
\label{eq:two-sided-equation}
\end{equation}
is equivalent after acting with \(A^{-1}\) to:
\begin{equation}
KB\phi=0.
\label{eq:two-sided-reduced}
\end{equation}
Therefore we find:
\begin{equation}
B\phi\in\ker K,
\label{eq:Bphi-kernel}
\end{equation}
and hence:
\begin{equation}
\ker K_{A,B}
=
B^{-1}\bigl(\ker K\bigr).
\label{eq:two-sided-kernel}
\end{equation}
Equation \eqref{eq:two-sided-kernel} means that the zero modes are mapped bijectively by the invertible field transformation \(B^{-1}\). Their number and linear structure are therefore preserved even though the explicit representatives of the zero modes may change.

In momentum space we can similarly obtain:
\begin{equation}
\det K_{A,B}(p)
=
\det A(p)\,
\det K(p)\,
\det B(p).
\label{eq:two-sided-det}
\end{equation}
If both \(A(p)\) and \(B(p)\) are nonsingular, the finite-momentum zero set of \(K_{A,B}(p)\) is identical to that of \(K(p)\).

A special case is a similarity transformation:
\begin{equation}
K_{\mathrm{sim}}
=
B^{-1}KB,
\label{eq:similarity-transformation}
\end{equation}
but unlike general left multiplication, Eq.~\eqref{eq:similarity-transformation} preserves the complete spectrum of \(K\), subject to the usual domain assumptions for unbounded operators:
\begin{equation}
\sigma\!\left(K_{\mathrm{sim}}\right)
=
\sigma(K),
\label{eq:spectrum-similarity}
\end{equation}
where \(\sigma(K)\) is the spectrum of \(K\). This is stronger than Eq.~\eqref{eq:kernel-invariance-general}, which guarantees only the preservation of zero modes.

This theorem applies to any free continuum field theory whose kinetic operator is multiplied by an injective operator. For a scalar field with local kinetic operator:
\begin{equation}
K_0(p)=p^2-m^2,
\label{eq:scalar-local-operator}
\end{equation}
where \(p^2=\eta^{\mu\nu}p_\mu p_\nu\), \(\eta^{\mu\nu}\) is the inverse spacetime metric, and \(m\) is the scalar mass, we can consider:
\begin{equation}
K_A(p)=A(p)\bigl(p^2-m^2\bigr).
\label{eq:scalar-deformed-operator}
\end{equation}
If \(A(p)\neq 0\) for every finite \(p\), then:
\begin{equation}
K_A(p)=0
\quad\Longleftrightarrow\quad
p^2=m^2,
\label{eq:scalar-shell-invariance}
\end{equation}
so the deformation therefore introduces no additional free scalar mass shells.

For a vector field, we let $K_{\mu\nu}(p)$ be the gauge-fixed kinetic matrix, where \(\mu\) and \(\nu\) are Lorentz indices. Then if:
\begin{equation}
K^{A}_{\mu\nu}(p)
=
A_{\mu}{}^{\rho}(p)K_{\rho\nu}(p),
\label{eq:vector-deformation}
\end{equation}
and the matrix \(A_{\mu}{}^{\rho}(p)\) is invertible, then the gauge-fixed zero set is unchanged. Gauge fixing, or equivalently restriction to the physical quotient space, is necessary because the ungauge-fixed operator already possesses zero modes generated by gauge redundancy.

The same reasoning applies to a gauge-fixed linearized gravitational operator:
\begin{equation}
K_{\mu\nu}{}^{\rho\sigma}(p),
\label{eq:gravity-kinetic}
\end{equation}
which acts on symmetric tensor perturbations \(h_{\rho\sigma}\). Multiplication by an invertible tensor-valued form factor cannot create additional free graviton poles, while it may modify the ultraviolet weighting of the propagator, but it cannot change its finite pole set unless the multiplying operator itself becomes singular or noninvertible.

The fermionic theorem proved is recovered by setting:
\begin{equation}
K=D_0=i\gamma^\mu D_\mu-m
\label{eq:general-to-Dirac}
\end{equation}
and:
\begin{equation}
A
=
F^{-1}\!\left(\frac{\Box_D}{E_M^2}\right),
\label{eq:general-to-nonlocal}
\end{equation}
where \(\gamma^\mu\) are the Dirac matrices, \(D_\mu\) is the gauge-covariant derivative, \(m\) is the fermion mass, \(\Box_D=D_\mu D^\mu\) is the covariant d'Alembertian, \(E_M>0\) is the nonlocal energy scale, and \(F\) is a zero-free entire function. The general identity \eqref{eq:kernel-invariance-general} then gives:
\begin{equation}
\ker\!\left[
F^{-1}\!\left(\frac{\Box_D}{E_M^2}\right)D_0
\right]
=
\ker D_0,
\label{eq:Dirac-as-corollary}
\end{equation}
the absence of fermion doubling in the continuum nonlocal Dirac theory is therefore a corollary of the more general invertible-deformation theorem.

Now we note that since the Nielsen–-Ninomiya theorem is usually stated for chiral fermions, it is important for us to mention that the same conclusion holds for Weyl fermions in the continuum nonlocal theory. The Dirac proof above already contains the massless case by setting \(m=0\), but the chiral statement is slightly sharper because it asks whether the nonlocal form factor can generate an additional mirror fermion of
opposite chirality.

We first define the usual chiral projection operators by:
\begin{equation}
P_L=\frac{1-\gamma^5}{2},
\qquad
P_R=\frac{1+\gamma^5}{2},
\end{equation}
where \(\gamma^5\) is the chirality matrix, \(P_L\) projects onto left-handed spinors, and \(P_R\) projects onto right-handed spinors. For a massless Dirac fermion the local kinetic operator is:
\begin{equation}
\mathcal D_0=i\gamma^\mu D_\mu,
\end{equation}
where \(D_\mu\) is the appropriate gauge-covariant derivative acting on the fermion representation~\cite{Dirac1928Electron,Weyl1929}. If the gauge connection does not mix left-handed and right-handed fields, then the covariant d'Alembertian \(\Box_D=D_\mu D^\mu\) preserves the chiral subspaces. So the scalar nonlocal form factor does not convert a left-handed fermion into a right-handed fermion, or a right-handed fermion into a left-handed fermion as it only multiplies the corresponding chiral kinetic operator by an invertible operator.

Equivalently we can note that in two-component notation, a free left-handed Weyl fermion has inverse propagator:
\begin{equation}
D_L(p)=p_\mu \bar{\sigma}^\mu,
\end{equation}
where \(p_\mu\) is the continuous four-momentum and \(\bar{\sigma}^\mu=(1,-\sigma^i)\), with \(\sigma^i\) the Pauli matrices~\cite{PeskinSchroeder1995,WeinbergQFTI1995,DreinerHaberMartin2010,Weyl1929,VanderWaerden1928,Pauli1927}. The corresponding continuum nonlocal Weyl operator is:
\begin{equation}
D_{F,L}(p)
=
F^{-1}\!\left(-\frac{p^2}{E_M^2}\right)
p_\mu \bar{\sigma}^\mu .
\end{equation}
And taking the determinant gives us:
\begin{equation}
\det D_{F,L}(p)
=
F^{-2}\!\left(-\frac{p^2}{E_M^2}\right)
\det\!\left(p_\mu \bar{\sigma}^\mu\right),
\end{equation}
and using:
\begin{equation}
\det\!\left(p_\mu \bar{\sigma}^\mu\right)=p^2,
\end{equation}
we get:
\begin{equation}
\det D_{F,L}(p)
=
F^{-2}\!\left(-\frac{p^2}{E_M^2}\right)p^2 .
\end{equation}
Now because the admissible form factor satisfies:
\begin{equation}
\notag F(z)\neq 0
\qquad
\forall\;\;\text{finite}\;\;z\in\mathbb C,
\end{equation}
the nonlocal prefactor is finite and nonzero at every finite momentum. So therefore:
\begin{equation}
\det D_{F,L}(p)=0
\qquad\Longleftrightarrow\qquad
p^2=0 .
\end{equation}
So the nonlocal left-handed Weyl operator has exactly the same finite-momentum zero set as the ordinary left-handed Weyl operator.

The same argument also applies to a right-handed Weyl fermion, whose free inverse propagator is:
\begin{equation}
D_R(p)=p_\mu \sigma^\mu,
\end{equation}
with \(\sigma^\mu=(1,\sigma^i)\).  The nonlocal right-handed operator is:
\begin{equation}
D_{F,R}(p)
=
F^{-1}\!\left(-\frac{p^2}{E_M^2}\right)
p_\mu \sigma^\mu ,
\end{equation}
and hence:
\begin{equation}
\det D_{F,R}(p)=0
\qquad\Longleftrightarrow\qquad
p^2=0 .
\end{equation}

The conclusion we can draw from this is that a zero-free entire-function deformation does not produce an additional Weyl node, an additional chiral species, or a mirror fermion. It preserves the chiral zero structure of the continuum theory. Fermion doubling in the Nielsen–-Ninomiya sense requires additional inequivalent zeros of the inverse propagator on the momentum manifold. Since the continuum nonlocal Weyl operator is obtained from the ordinary Weyl operator by multiplication with an invertible nonzero function, no such additional chiral zeros are
created.

A corresponding statement holds for nonlinear field equations, we would want to look at these because most physically complete field theories are nonlinear once interactions are included~\cite{PeskinSchroeder1995,WeinbergQFTI1995}. For example Yang–Mills theory, general relativity, and interacting scalar theories all have nonlinear equations of motion, so looking at nonlinear equations would show whether the same invertibility principle preserves not merely free zero modes, but the full classical solution set. To see this we can let:
\begin{equation}
\mathcal{E}[\phi]=0
\label{eq:nonlinear-equation}
\end{equation}
be an equation of motion for a field configuration \(\phi\), where \(\mathcal{E}\) is not assumed to be linear. Then we let \(A[\phi]\) be an operator that is injective for every admissible field configuration, and we can define:
\begin{equation}
\mathcal{E}_A[\phi]
=
A[\phi]\mathcal{E}[\phi].
\label{eq:nonlinear-deformation}
\end{equation}
Then we see:
\begin{equation}
\mathcal{E}_A[\phi]=0
\quad\Longleftrightarrow\quad
\mathcal{E}[\phi]=0,
\label{eq:nonlinear-solution-equivalence}
\end{equation}
so the two equations have the same classical solution set, now this statement does not by itself establish quantum equivalence, because the functional measure, interaction vertices, Jacobians, boundary conditions, and gauge identities may differ.

For an interacting quantum field theory the exact inverse propagator may be written as:
\begin{equation}
\Gamma^{(2)}_A(p)
=
A(p)K_0(p)-\Sigma_A(p),
\label{eq:interacting-inverse}
\end{equation}
where \(K_0(p)\) is the free local kinetic operator, \(A(p)\) is the deforming multiplier, \(\Sigma_A(p)\) is the interacting self-energy, and \(\Gamma^{(2)}_A(p)\) is the exact one-particle-irreducible two-point function. The physical poles are determined by:
\begin{equation}
\det\Gamma^{(2)}_A(p)=0.
\label{eq:interacting-poles}
\end{equation}
Note that the free kernel theorem does not imply that Eq.~\eqref{eq:interacting-poles} has exactly the same solutions as the undeformed interacting theory, because the self-energy need not factor through \(A(p)\).

The interacting pole set is preserved only if the exact two-point function admits a factorization of the form:
\begin{equation}
\Gamma^{(2)}_A(p)
=
A_{\mathrm{eff}}(p)\Gamma^{(2)}_0(p),
\label{eq:exact-factorization}
\end{equation}
where \(A_{\mathrm{eff}}(p)\) is finite and invertible and \(\Gamma^{(2)}_0(p)\) is the exact inverse propagator of the comparison theory. Under an additional condition, that the physical particle spectrum remains invariant under background deformations, allowing for a clean decoupling of vacuum dynamics from background field interactions, we write this as:
\begin{equation}
\det\Gamma^{(2)}_A(p)
=
\det A_{\mathrm{eff}}(p)\,
\det\Gamma^{(2)}_0(p),
\label{eq:exact-det-factorization}
\end{equation}
and the two exact propagators possess the same finite pole set. Without Eq.~\eqref{eq:exact-factorization}, interactions may shift particle masses or generate bound-state, resonance, or instability poles. Such dynamically generated structures are different from potential regulator-induced doubling in the free kinetic operator.

The general conclusion is that an injective deformation of a linear kinetic operator preserves its kernel, and an everywhere nonsingular momentum-space multiplier preserves its finite zero set. This result applies to scalar, fermionic, vector, gravitational, and other continuum systems. The absence of fermion doubling in the nonlocal Dirac theory is one particular application of this general operator-theoretic principle.

\section{A General Criterion for Fermion Doubling}

We now can write down a set of criteria that can be applied to any arbitrary translation-invariant quadratic fermion theory, this result does not assume that spacetime is either continuous or discrete, nor does it assume locality or nonlocality.  It helps us determine the number of fermion species directly from the local structure of the inverse propagator in momentum space.

The theorem below is a statement about the exact continuum operator, and not about arbitrary finite truncations or arbitrary later lattice discretizations. We assume that the form factor is entire and nonvanishing at every finite argument, that $F(\Box_D/E_M^2)$ is defined on a common invariant domain for the fields you are studying, and that the inverse operator exists on that domain. In a general background the theorem proves equality of kernels, in the translation-invariant vacuum this becomes equality of the finite-momentum zero sets and thus equality of the free one-particle pole locations.

We first let $\mathcal{M}$ be the momentum manifold of the theory.  For a continuum theory we ordinarily would have $\mathcal{M}=\mathbb{R}^{d}$, whereas for a lattice theory $\mathcal{M}$ is a compact Brillouin torus, we then allow $\mathcal{M}$ to possess exact momentum identifications, written as:
\begin{equation}
p\sim p',
\end{equation}
meaning that $p$ and $p'$ label the same physical momentum mode.

We then let $V$ be the finite-dimensional complex vector space carrying the spinor, flavour, and any other internal indices of the fermion field. The quadratic Euclidean action is then written as:
\begin{equation}
S_{2}
=
\int_{\mathcal{M}}
\frac{d^{d}p}{(2\pi)^{d}}\,
\overline{\psi}(-p)\,
\mathscr{D}(p)\,
\psi(p),
\label{eq:general-quadratic-action}
\end{equation}
where:
\begin{equation}
\mathscr{D}(p):V\longrightarrow V
\end{equation}
is the momentum-space inverse fermion propagator.  We also assume that $\mathscr{D}(p)$ is continuously differentiable near each of its finite-momentum zeros.

The singular set of the inverse propagator is given by:
\begin{equation}
\mathcal{Z}
:=
\left\{
p\in\mathcal{M}:
\det\mathscr{D}(p)=0
\right\}.
\label{eq:fermionic-zero-set}
\end{equation}
We note that only inequivalent elements of the quotient $\mathcal{Z}/\!\sim$ are to be counted, since points related by an exact momentum identification represent the same physical mode.

A zero $p^{(i)}\in\mathcal{Z}$ is called a Dirac-regular zero of multiplicity $m_i$ if there exist a
neighborhood of $p^{(i)}$, smooth invertible matrices $L_i(q)$ and
$R_i(q)$, and a decomposition of the internal space such that for
$p=p^{(i)}+q$:
\begin{widetext}
\begin{equation}
\begin{aligned}
\begin{split}
L_i(q)\,
\mathscr{D}\!\left(p^{(i)}+q\right)
R_i(q)
=
\begin{pmatrix}
i\gamma^{a}e^{\ \mu}_{i\,a}q_{\mu}
 \otimes \mathbf{1}_{m_i}
 +O(|q|^{2})
&
O(|q|)
\\[4pt]
O(|q|)
&
H_i(q)
\end{pmatrix},
\label{eq:Dirac-regular-normal-form}
\end{split}
\end{aligned}
\end{equation}
\end{widetext}
where $\gamma^{a}$ are gamma matrices satisfying the $d$-dimensional Euclidean Clifford algebra like before:
\begin{equation}
\notag \{\gamma^{a},\gamma^{b}\}
=
2\delta^{ab}\mathbf{1},
\end{equation}
$\mathbf{1}_{m_i}$ is the identity on an $m_i$-dimensional flavour
space, $e^{\ \mu}_{i\,a}$ is a nonsingular matrix where:
\begin{equation}
\det\!\left(e^{\ \mu}_{i\,a}\right)\neq 0,
\end{equation}
and the heavy block $H_i(q)$ is invertible throughout a sufficiently
small neighborhood of $q=0$.

The integer $m_i$ measures the number of independent fermion species whose low-energy excitations are centered at $p^{(i)}$.  The nonsingularity of $e^{\ \mu}_{i\,a}$ ensures that the dispersion is linear in every momentum direction and therefore that the zero describes an ordinary relativistic Dirac mode rather than a higher-order or degenerate excitation.

\begin{theorem}[General fermion-species counting criterion]
\label{thm:general-doubling-criterion}
Let $\mathscr{D}(p)$ be the inverse propagator of a translation-invariant quadratic fermion theory on the momentum manifold $\mathcal{M}$. And suppose that:

\begin{enumerate}
\item the quotient zero set $\mathcal{Z}/\!\sim$ contains finitely many isolated classes
\begin{equation}
[p^{(1)}],\ldots,[p^{(N)}];
\end{equation}

\item every zero class is Dirac-regular, with multiplicity $m_i$;

\item $\mathscr{D}(p)$ is invertible away from these zero classes.
\end{enumerate}

Then the exact number of massless Dirac fermion species described by
the quadratic theory is
\begin{equation}
N_{\mathrm{species}}
=
\sum_{i=1}^{N}m_i.
\label{eq:species-number}
\end{equation}

If the intended continuum theory contains $N_{\mathrm{target}}$ fermion species, then the theory is free of fermion doubling if and only if
\begin{equation}
N_{\mathrm{species}}
=
N_{\mathrm{target}}.
\label{eq:no-doubling-condition}
\end{equation}
It possesses additional fermion species, that are conventionally called doublers, if
\begin{equation}
N_{\mathrm{species}}
>
N_{\mathrm{target}},
\end{equation}
and the number of additional species is
\begin{equation}
N_{\mathrm{doubler}}
=
N_{\mathrm{species}}
-
N_{\mathrm{target}}.
\label{eq:number-of-doublers2}
\end{equation}
\end{theorem}

The proof starts by choosing one representative $p^{(i)}$ from each inequivalent zero class. Since the zero classes are isolated, mutually disjoint neighborhoods may be chosen around them.

Inside the neighborhood of $p^{(i)}$, the smooth invertible transformations $L_i(q)$ and $R_i(q)$ do not change the rank of the kinetic operator, the zero set of its determinant, or the number of
propagating modes. Equation
\eqref{eq:Dirac-regular-normal-form} therefore gives us a locally equivalent description of the theory.

We write the transformed operator in block form as:
\begin{equation}
\mathscr{D}_{i}(q)
=
\begin{pmatrix}
A_i(q)&B_i(q)\\
C_i(q)&H_i(q)
\end{pmatrix},
\end{equation}
where $H_i(q)$ is invertible.  Smooth block Gaussian elimination gives:
\begin{equation}
\begin{pmatrix}
\mathbf{1}&-B_iH_i^{-1}\\
0&\mathbf{1}
\end{pmatrix}
\mathscr{D}_{i}(q)
\begin{pmatrix}
\mathbf{1}&0\\
-H_i^{-1}C_i&\mathbf{1}
\end{pmatrix}
=
\begin{pmatrix}
S_i(q)&0\\
0&H_i(q)
\end{pmatrix},
\label{eq:schur-reduction}
\end{equation}
where:
\begin{equation}
S_i(q)
=
A_i(q)-B_i(q)H_i(q)^{-1}C_i(q)
\end{equation}
is the Schur complement.

Because $B_i(q)=O(|q|)$ and $C_i(q)=O(|q|)$, their contribution to the Schur complement begins at second order. Consequently we find that:
\begin{equation}
S_i(q)
=
i\gamma^{a}e^{\ \mu}_{i\,a}q_{\mu}
 \otimes\mathbf{1}_{m_i}
+
O(|q|^{2}).
\label{eq:local-effective-Dirac}
\end{equation}
Since $e^{\ \mu}_{i\,a}$ is nonsingular, the linear change of momentum coordinates:
\begin{equation}
k_a=e^{\ \mu}_{i\,a}q_\mu,
\end{equation}
reduces the leading term to:
\begin{equation}
S_i(q)
=
i\gamma^{a}k_a\otimes\mathbf{1}_{m_i}
+
O(|k|^{2}).
\end{equation}
This is $m_i$ independent copies of the continuum Dirac
operator. The block $H_i(q)$ remains invertible and hence produces no additional gapless fermion.  So the zero class
$[p^{(i)}]$ contributes exactly $m_i$ fermion species.

Distinct zero classes possess disjoint momentum neighborhoods and are not related by an exact momentum identification.  Their low-energy fields are therefore independent. Summing their
multiplicities gives us:
\begin{equation}
N_{\mathrm{species}}
=
\sum_i m_i.
\end{equation}

On the other hand a massless fermionic excitation requires the propagator $\mathscr{D}(p)^{-1}$ to become singular, this can occur only where $\mathscr{D}(p)$ is noninvertible, precisely at a point of $\mathcal{Z}$. By hypothesis all such points have already been included and are Dirac-regular.  The above list thus exhausts all massless fermion species.  Comparing their total number with the intended number $N_{\mathrm{target}}$ proves Eqs.~\eqref{eq:no-doubling-condition} and \eqref{eq:number-of-doublers2}.

For a broad class of massless fermion theories the inverse propagator has the form:
\begin{equation}
\mathscr{D}(p)
=
i\sum_{a=1}^{d}\gamma^{a}f_a(p),
\label{eq:gamma-linear-operator}
\end{equation}
where the functions $f_a(p)$ are real on the Euclidean momentum
domain. In this case we have:
\begin{equation}
\mathscr{D}(p)=0
\quad\Longleftrightarrow\quad
f_a(p)=0
\quad\forall a.
\end{equation}

We let $p^{(i)}$ be a simultaneous zero and define its Jacobian by:
\begin{equation}
J^{(i)}_{a\mu}
:=
\left.
\frac{\partial f_a(p)}{\partial p_\mu}
\right|_{p=p^{(i)}}.
\label{eq:jacobian-definition}
\end{equation}
A Taylor expansion gives us:
\begin{equation}
f_a\!\left(p^{(i)}+q\right)
=
J^{(i)}_{a\mu}q_\mu
+
O(|q|^{2}).
\end{equation}
If:
\begin{equation}
\det J^{(i)}\neq 0,
\label{eq:regular-jacobian}
\end{equation}
then $p^{(i)}$ is a Dirac-regular zero of multiplicity one.  Therefore for an operator of the form \eqref{eq:gamma-linear-operator}:
\begin{equation}
N_{\mathrm{species}}
=
\#\left\{
[p]\in\mathcal{Z}/\!\sim:
\det J(p)\neq 0
\right\},
\label{eq:practical-species-count}
\end{equation}
so long as all zeros are regular.

For a chiral theory each such zero also carries the orientation index:
\begin{equation}
\nu_i
=
\operatorname{sgn}\det J^{(i)}.
\label{eq:zero-chirality-index}
\end{equation}
The number of zeros determines the number of species, while the signs $\nu_i$ determine their relative chiral orientations. On a compact Brillouin torus, the assumptions of the Nielsen–-Ninomiya theorem force these indices to occur in compensating combinations. The species-counting theorem itself however requires neither compact
momentum space nor locality.

The criterion gives a general no-doubling result for invertible deformations.

\begin{corollary}[Invariance of the fermion count]
\label{cor:invertible-species-invariance}
Let
\begin{equation}
\mathscr{D}'(p)
=
A(p)\mathscr{D}(p)B(p),
\label{eq:left-right-deformation}
\end{equation}
where $A(p)$ and $B(p)$ are smooth and invertible for every finite
$p\in\mathcal{M}$.  Then $\mathscr{D}'$ and $\mathscr{D}$ have:

\begin{enumerate}
\item the same finite-momentum zero set;
\item the same nullity at every momentum;
\item the same Dirac multiplicity at every regular zero;
\item the same total number of fermion species.
\end{enumerate}
This shows that an everywhere-invertible deformation cannot introduce fermion doublers.
\end{corollary}

The proof appears as at every momentum:
\begin{equation}
\det\mathscr{D}'(p)
=
\det A(p)\,
\det\mathscr{D}(p)\,
\det B(p).
\end{equation}
Since $\det A(p)$ and $\det B(p)$ are nonzero, we have:
\begin{equation}
\det\mathscr{D}'(p)=0
\quad\Longleftrightarrow\quad
\det\mathscr{D}(p)=0.
\end{equation}
Moreover we have:
\begin{equation}
\ker\mathscr{D}'(p)
=
B(p)^{-1}\ker\mathscr{D}(p),
\end{equation}
so the kernels are isomorphic and have equal dimension. Smooth invertible left and right multiplication are precisely the equivalences permitted in the Dirac-regular normal form \eqref{eq:Dirac-regular-normal-form}, so they therefore preserve each multiplicity $m_i$. The species-counting formula then gives the same total fermion number for both operators.

For the continuum nonlocal Dirac theory we study:
\begin{equation}
\notag \mathscr{D}_{F}(p)
=
F^{-1}\!\left(-\frac{p^{2}}{E_M^{2}}\right)
\mathscr{D}_{0}(p),
\end{equation}
so that:
\begin{equation}
\notag A(p)
=
F^{-1}\!\left(-\frac{p^{2}}{E_M^{2}}\right)\mathbf{1},
\qquad
B(p)=\mathbf{1}.
\end{equation}
If $F(z)$ has no finite zeros, then $A(p)$ is invertible at every finite momentum.  The corollary~\ref{cor:invertible-species-invariance} shows not only that the zero set is unchanged, but also that the multiplicity and local Dirac structure of every zero are unchanged in the theory. So the complete nonlocal theory contains exactly the same number of fermion species as its local parent theory.

A finite polynomial truncation of the form factor need not satisfy this hypothesis, this comment will become important soon in the paper. If the truncated multiplier develops a zero or a pole, then invertibility fails and the deformation must be tested again using Theorem~\ref{thm:general-doubling-criterion}.

\section{A Test for Genuine and False Doubling}

Now that we have given the criteria for doubling to occur and before we explain if it is physical or mathematical, we want to give a simple test for deciding whether the apparent extra fermion mode is a genuine doubler or only a false artifact of the theory. First we let \(D_{\rm exact}(p)\) be the exact inverse fermion propagator of the theory, where \(p\) is the momentum variable on the physical momentum space \(\mathcal M\). The possible fermionic modes are determined by the zero set by:
\begin{equation}
    Z_{\rm exact}
    =
    \left\{
    p\in \mathcal M :
    \det D_{\rm exact}(p)=0
    \right\},
\end{equation}
there are two zeros, \(p\) and \(p'\), they are counted as the same physical mode if they are related by an exact momentum identification \(p\sim p'\).

An apparent extra fermion is a genuine doubler only if there exists an inequivalent zero:
\begin{equation}
    p_* \in Z_{\rm exact},
    \qquad
    p_* \not\sim p_{\rm phys},
\end{equation}
where \(p_{\rm phys}\) is the ordinary Dirac mass-shell zero, and if the expansion of \(D_{\rm exact}(p)\) around \(p_*\) contains an independent Dirac cone:
\begin{equation}
    D_{\rm exact}(p_*+q)
    =
    i\Gamma^\mu q_\mu
    +
    O(q^2),
    \qquad
    \det(\Gamma^\mu q_\mu)\not\equiv 0 ,
\end{equation}
where \(q_\mu\) is the small momentum measured relative to \(p_*\), and \(\Gamma^\mu\) are the effective gamma matrices governing the low-energy excitation near that zero.

But the apparent mode is a false doubler if it occurs only after replacing the exact nonlocal operator by a finite polynomial truncation:
\begin{equation}
    D_N(p)
    =
    F_N^{-1}\!\left(-\frac{p^2}{E_M^2}\right)(\slashed p-m),
\end{equation}
where \(F_N\) is a finite Taylor approximation to the full entire function \(F\).  If:
\begin{equation}
    \det D_N(p_*)=0
    \qquad
    \text{but}
    \qquad
    \det D_{\rm exact}(p_*)\neq 0,
\end{equation}
then \(p_*\) is not a physical fermion species of the nonlocal theory. It is a truncation artifact produced by replacing the exact zero-free entire function with a different finite-order differential operator.

So the test is to count only inequivalent zeros of the exact inverse propagator. Zeros present in a finite truncation but absent in the full zero-free entire-function theory are false doublers, not genuine Nielsen–-Ninomiya fermion doublers.

\section{A TEST PROCEDURE FOR MATHEMATICAL AND PHYSICAL FERMION DOUBLING}
\label{sec:physical-mathematical-doubling-test}

Now the existence of a zero of a fermionic kinetic operator is a necessary condition for an additional fermionic excitation, but that in itself is not good enough to show the existence of an additional physical particle. So we believe that it is useful to distinguish three logically different statements. The first is an algebraic zero, a Dirac-regular zero, and a physical fermion state. Neither implication in the three-level-hierarchy is automatic, since a repeated algebraic zero may show up from spinor degeneracy or from a finite-order truncation (this will be discussed later), while a Dirac-like pole may correspond to a
ghost, tachyon, resonance, or and otherwise nonphysical state. We now can give a procedure for distinguishing these possibilities.

We start with the exact free inverse propagator, let $\mathcal{M}$ represent the momentum manifold of the theory. For an ordinary continuum theory we take:
\begin{equation}
\mathcal{M}=\mathbb{R}^{d},
\end{equation}
where $d$ is the number of spacetime dimensions, while for a lattice theory $\mathcal{M}$ is a compact Brillouin torus. We allow the momentum manifold to possess exact identifications:
\begin{equation}
p\sim p',
\label{eq:momentum-equivalence-physical-test}
\end{equation}
where $p$ and $p'$ represent the same physical momentum mode. For
example lattice momenta that differ by a reciprocal-lattice vector are equivalent under Eq.~\eqref{eq:momentum-equivalence-physical-test}.

Now let $V$ be the finite-dimensional complex vector space carrying all spinor, flavour, and internal indices of the fermion field. The exact free quadratic action is written as:
\begin{equation}
S_{2}
=
\int_{\mathcal{M}}
\frac{d^{d}p}{(2\pi)^{d}}\,
\overline{\psi}(-p)\,
D_{\mathrm{ex}}(p)\,
\psi(p),
\label{eq:exact-free-quadratic-action}
\end{equation}
where $\psi(p)\in V$ is the momentum-space fermion field, $\overline{\psi}(p)$ is its Dirac adjoint, and:
\begin{equation}
D_{\mathrm{ex}}(p):V\longrightarrow V
\label{eq:exact-free-inverse-propagator-map}
\end{equation}
is the exact free inverse propagator. The subscript ``ex'' just means that the complete kinetic operator is being used. In particular we do not have $D_{\mathrm{ex}}$ as a finite Taylor or derivative truncation of
the theory as this would spoil the results.

At momenta where $D_{\mathrm{ex}}(p)$ is invertible, the exact free
propagator is given by:
\begin{equation}
S_{\mathrm{ex}}(p)
=
iD_{\mathrm{ex}}(p)^{-1}.
\label{eq:exact-free-propagator}
\end{equation}
The first step in every doubling test must be to determine $D_{\mathrm{ex}}(p)$ and $S_{\mathrm{ex}}(p)$ before making any approximation.

Step one in our procedure is to determine the exact zero set.

The finite-momentum singular set of the exact free inverse propagator is:
\begin{equation}
\mathcal{Z}_{\mathrm{ex}}
=
\left\{
p\in\mathcal{M}
\ \middle|\
\det D_{\mathrm{ex}}(p)=0
\right\},
\label{eq:exact-singular-set-test}
\end{equation}
where $\det D_{\mathrm{ex}}(p)$ is the determinant over the finite internal vector space $V$. Equation~\eqref{eq:exact-singular-set-test} identifies momenta at which the kinetic matrix has a nontrivial kernel:
\begin{equation}
\det D_{\mathrm{ex}}(p)=0
\quad\Longleftrightarrow\quad
\ker D_{\mathrm{ex}}(p)\neq\{0\}.
\label{eq:det-kernel-equivalence-test}
\end{equation}

Only inequivalent momentum points are to be counted and thus the relevant set is the quotient:
\begin{equation}
\widehat{\mathcal{Z}}_{\mathrm{ex}}
=
\mathcal{Z}_{\mathrm{ex}}/\!\sim,
\label{eq:quotient-zero-set-test}
\end{equation}
where the equivalence relation $\sim$ is the one defined in Eq.~\eqref{eq:momentum-equivalence-physical-test}. Two representatives
of the same equivalence class:
\begin{equation}
[p]
=
\left\{
p'\in\mathcal{M}:p'\sim p
\right\}
\label{eq:momentum-equivalence-class}
\end{equation}
describe one momentum mode and cannot be counted as two fermion
species.

A large algebraic multiplicity of
$\det D_{\mathrm{ex}}(p)$ can also not be confused with a large number
of species. For example in four dimensions,:
\begin{equation}
\det(\slashed{p}-m)
=
\left(p^{2}-m^{2}\right)^{2},
\label{eq:ordinary-Dirac-determinant}
\end{equation}
where:
\begin{equation}
\slashed{p}=\gamma^{\mu}p_{\mu},
\qquad
p^{2}=\eta^{\mu\nu}p_{\mu}p_{\nu},
\end{equation}
the matrices $\gamma^{\mu}$ are the Dirac gamma matrices and $\eta^{\mu\nu}$ is the inverse Minkowski metric. The square in Eq.~\eqref{eq:ordinary-Dirac-determinant} reflects ordinary spinor degeneracy. It does not mean that a single Dirac field represents two independent fermion species.

Step two is to test whether each zero is Dirac-regular.

We first let $p^{(j)}$ be a representative of an inequivalent class $[p^{(j)}]\in\widehat{\mathcal{Z}}_{\mathrm{ex}}$, and introduce a
small momentum displacement:
\begin{equation}
p=p^{(j)}+q,
\qquad
|q|\ll 1,
\label{eq:momentum-expansion-around-candidate}
\end{equation}
where the vector $q$ is the momentum measured relative to the candidate zero $p^{(j)}$.

The zero $p^{(j)}$ is called a Dirac-regular zero of multiplicity $m_{j}$ if there exist smooth invertible matrices $L_{j}(q)$ and $R_{j}(q)$, a decomposition of the internal space $V$, and a neighbourhood of $q=0$ in which:
\begin{widetext}
\begin{equation}
\begin{aligned}
\begin{split}
L_{j}(q)
D_{\mathrm{ex}}\!\left(p^{(j)}+q\right)
R_{j}(q)
=
\begin{pmatrix}
i\gamma^{a}e^{\mu}_{ja}q_{\mu}
 \otimes\mathbf{1}_{m_{j}}
 +O(|q|^{2})
&
O(|q|)
\\[2mm]
O(|q|)
&
H_{j}(q)
\end{pmatrix},
\label{eq:Dirac-regular-normal-form-test}
\end{split}
\end{aligned}
\end{equation}
\end{widetext}
here $a$ is a local orthonormal spacetime index, $\mu$ is a momentum-coordinate index, and the matrices $\gamma^{a}$ satisfy:
\begin{equation}
\notag \{\gamma^{a},\gamma^{b}\}
=
2\delta^{ab}\mathbf{1}.
\label{eq:local-clifford-algebra-test}
\end{equation}
The matrix $e^{\mu}_{ja}$ determines the linear relation between the local momentum coordinates and the Dirac coefficients. It also must be nonsingular:
\begin{equation}
\det\!\left(e^{\mu}_{ja}\right)\neq 0,
\label{eq:nonsingular-Dirac-velocity-matrix}
\end{equation}
the symbol $\mathbf{1}_{m_{j}}$ is the identity matrix on an $m_{j}$-dimensional flavour or species space. The integer $m_{j}$ is therefore the number of independent Dirac species associated with the zero $p^{(j)}$.

The matrix $H_{j}(q)$ is the block acting on all remaining, nonzero modes. It must remain invertible in a sufficiently small neighbourhood:
\begin{equation}
\det H_{j}(q)\neq 0.
\label{eq:heavy-block-invertibility}
\end{equation}
The terms represented by $O(|q|)$ vanish at least linearly as $q\to0$, while the terms denoted by $O(|q|^{2})$ vanish at least quadratically.

The invertible matrices $L_{j}$ and $R_{j}$ represent nonsingular changes of basis so they cannot create or remove zero modes. So Eq.~\eqref{eq:Dirac-regular-normal-form-test} says that the candidate zero locally contains $m_{j}$ independent copies of the ordinary Dirac operator.

A zero that fails the test \eqref{eq:Dirac-regular-normal-form-test} is not automatically a fermion doubler as it may instead be a higher-order algebraic degeneracy, a constrained mode, a removable singularity, or a non-Dirac higher-derivative excitation, so we need not panic yet.

The total mathematical fermion count isL
\begin{equation}
N_{\mathrm{math}}
=
\sum_{[p^{(j)}]\in
\widehat{\mathcal{Z}}_{\mathrm{ex}}^{\,\mathrm{D}}}
m_{j},
\label{eq:mathematical-species-count}
\end{equation}
where $\widehat{\mathcal{Z}}_{\mathrm{ex}}^{\,\mathrm{D}}$ is the subset of inequivalent zeros that satisfy the Dirac-regularity condition. If the intended theory contains $N_{\mathrm{int}}$ fermion species, then mathematical or spectral doubling occurs when:
\begin{equation}
\notag N_{\mathrm{math}}>N_{\mathrm{int}}.
\label{eq:mathematical-doubling-condition}
\end{equation}
So mathematical doubling just means the existence of additional inequivalent Dirac-regular zeros of the exact free inverse propagator.

Step three is to test for a nonremovable propagator pole.

Dirac-regularity is a local statement about the kinetic operator, to establish a particle interpretation, one must next examine the propagator~\cite{PeskinSchroeder1995,WeinbergQFTI1995,StreaterWightman2000}.

After analytic continuation to Minkowski momentum, a simple fermion pole of mass $m_{j}$ has the local form:
\begin{equation}
S_{\mathrm{ex}}(p)
=
\frac{
 iZ_{j}
 \left(\slashed{p}+m_{j}\right)P_{j}
}{
 p^{2}-m_{j}^{2}+i\epsilon
}
+
S_{\mathrm{reg},j}(p),
\label{eq:physical-fermion-pole-form}
\end{equation}
where $m_{j}$ is the candidate physical mass, $Z_{j}$ is the wave-function residue, $P_{j}$ is a projector onto the corresponding flavour or internal subspace, and $S_{\mathrm{reg},j}(p)$ is regular at $p^{2}=m_{j}^{2}$. And we note that the positive quantity $\epsilon$ specifies the usual Feynman boundary condition.

The multiplicity of this pole is:
\begin{equation}
m_{j}
=
\operatorname{rank}P_{j}.
\label{eq:pole-projector-rank}
\end{equation}
The pole is nonremovable only if:
\begin{equation}
Z_{j}\neq 0,
\qquad
P_{j}\neq 0.
\label{eq:nonremovable-pole-condition}
\end{equation}
If the numerator vanishes in this way that the apparent denominator is cancelled, then the propagator is regular and no particle state is
present.

More generally we can say that the residue operator at the pole may be defined by:
\begin{equation}
\mathcal{R}_{j}(p)
=
\lim_{p^{2}\to m_{j}^{2}}
\left(p^{2}-m_{j}^{2}\right)
\left[-iS_{\mathrm{ex}}(p)\right].
\label{eq:general-residue-operator}
\end{equation}
For the simple Dirac pole
\eqref{eq:physical-fermion-pole-form}:
\begin{equation}
\mathcal{R}_{j}(p)
=
Z_{j}\left(\slashed{p}+m_{j}\right)P_{j}
\qquad
\text{at }p^{2}=m_{j}^{2}.
\label{eq:Dirac-residue-operator}
\end{equation}
A vanishing residue operator means that an apparent pole does not
create an asymptotic one-particle state.

Step four is to impose the physical-state conditions.

A mathematical Dirac pole represents a healthy physical fermion only if all of the following conditions are satisfied.

First, the mass shell must be real:
\begin{equation}
m_{j}^{2}\in\mathbb{R},
\qquad
m_{j}^{2}\geq 0.
\label{eq:real-mass-shell-condition}
\end{equation}
The corresponding positive-energy branch is:
\begin{equation}
p^{0}
=
+\sqrt{\boldsymbol{p}^{\,2}+m_{j}^{2}},
\label{eq:positive-energy-branch}
\end{equation}
where $\boldsymbol{p}$ is the spatial momentum. A negative value $m_{j}^{2}<0$ describes a tachyonic instability rather than a healthy fermion particle. A genuinely complex value of $m_{j}^{2}$ corresponds to a resonance, instability, or complex pole rather than a stable asymptotic particle~\cite{Barnett2023}.

Second, the pole must have nonzero residue, as required by Eq.~\eqref{eq:nonremovable-pole-condition}.

Third, the residue must have the physical sign. For a simple isolated Dirac pole in a unitary theory, this requires:
\begin{equation}
Z_{j}>0.
\label{eq:positive-residue-condition}
\end{equation}
Note that pole with $Z_{j}<0$ carries negative spectral weight and represents a ghost rather than a positive-norm physical fermion~\cite{StreaterWightman2000,PeskinSchroeder1995,WeinbergQFTI1995}.

In a theory with several coupled fermion fields, the scalar condition  given by \eqref{eq:positive-residue-condition} is replaced by positivity of the residue matrix on the physical one-particle subspace:
\begin{equation}
v^{\dagger}\mathcal{Z}_{j}v>0
\qquad
\forall\;\;\text{nonzero}
\;\;v\in\operatorname{Ran}P_{j},
\label{eq:positive-residue-matrix-condition}
\end{equation}
where $\mathcal{Z}_{j}$ is the wave-function residue matrix and $\operatorname{Ran}P_{j}$ is the range of the pole projector.

Fourth, the state must survive all constraints and projections defining
the physical Hilbert space, it must not be a gauge mode, a nonnormalizable solution, or a state removed by an exact constraint.

We define the physical admissibility indicator by:
\begin{equation}
\chi_{j}
=
\begin{cases}
1,
&
\text{if the candidate satisfies all physical-state tests},
\\[1mm]
0,
&
\text{otherwise}.
\end{cases}
\label{eq:physical-admissibility-indicator}
\end{equation}
The number of healthy physical fermion species is then:
\begin{equation}
N_{\mathrm{phys}}
=
\sum_{[p^{(j)}]\in
\widehat{\mathcal{Z}}_{\mathrm{ex}}^{\,\mathrm{D}}}
m_{j}\chi_{j}.
\label{eq:physical-species-count}
\end{equation}
Physical fermion doubling occurs when:
\begin{equation}
N_{\mathrm{phys}}>N_{\mathrm{int}}.
\label{eq:physical-doubling-condition}
\end{equation}

Equations~\eqref{eq:mathematical-doubling-condition} and \eqref{eq:physical-doubling-condition} make the distinction precise. What we call mathematical doubling counts additional inequivalent Dirac-regular zeros, while physical doubling counts only those zeros that also generate healthy positive-norm asymptotic fermions.

Step five is to distinguish doubling from dynamical poles.

For an interacting theory the exact one-particle-irreducible two-point function can be written as:
\begin{equation}
\Gamma^{(2)}(p)
=
D_{\mathrm{ex}}(p)-\Sigma(p),
\label{eq:interacting-exact-two-point-test}
\end{equation}
where $D_{\mathrm{ex}}(p)$ is the exact free inverse propagator and $\Sigma(p)$ is the fermion self-energy generated by interactions. The exact interacting propagator is given by:
\begin{equation}
S_{\mathrm{int}}(p)
=
i\left[\Gamma^{(2)}(p)\right]^{-1}.
\label{eq:interacting-propagator-test}
\end{equation}
Its poles will satisfy:
\begin{equation}
\det\Gamma^{(2)}(p)=0.
\label{eq:interacting-pole-equation-test}
\end{equation}

An additional Dirac-regular zero already present when:
\begin{equation}
\Sigma(p)=0
\label{eq:free-limit-self-energy}
\end{equation}
is a candidate regulator-induced or discretization-induced doubler. By contrast a pole that appears only after the self-energy is included may be a bound state, resonance, collective excitation, or interaction-induced instability. Such a pole is dynamically generated and is not fermion doubling in the sense of Nielsen–-Ninomiya~\cite{NielsenNinomiyaNoGo1981,NielsenNinomiyaI1981,NielsenNinomiyaII1981}.

Step six is to exclude finite-truncation artifacts.

For a moment suppose the exact nonlocal form factor is the zero-free entire function:
\begin{equation}
F(z)
=
\sum_{n=0}^{\infty}c_{n}z^{n},
\qquad
F(z)\neq0
\quad
\forall \;\;z\in\mathbb{C},
\label{eq:exact-zero-free-entire-test}
\end{equation}
where $c_{n}$ are the Taylor coefficients. Its finite Taylor
polynomial is then:
\begin{equation}
F_{N}(z)
=
\sum_{n=0}^{N}c_{n}z^{n},
\label{eq:finite-taylor-polynomial-test}
\end{equation}
where $N$ is a fixed nonnegative integer.

Because $F$ is entire its Taylor polynomials converge uniformly to $F$ on every compact subset $K\subset\mathbb{C}$:
\begin{equation}
\sup_{z\in K}
\left|F_{N}(z)-F(z)\right|
\longrightarrow0
\qquad
\text{as }N\to\infty.
\label{eq:compact-uniform-convergence-test}
\end{equation}
Suppose that $z_{N}$ is a root of the truncated form factor:
\begin{equation}
F_{N}(z_{N})=0.
\label{eq:truncated-root-test}
\end{equation}
If a subsequence of roots converged to a finite limit $z_{\ast}\in\mathbb{C}$, then:
\begin{equation}
z_{N}\longrightarrow z_{\ast}
\quad\Longrightarrow\quad
F(z_{\ast})
=
\lim_{N\to\infty}F_{N}(z_{N})
=
0.
\label{eq:truncation-root-limit}
\end{equation}
This contradicts the zero-free condition in Eq.~\eqref{eq:exact-zero-free-entire-test}. So roots of the Taylor polynomials cannot converge to a finite zero of the exact form factor. Any persistent sequence of truncation roots must leave every compact subset of the complex plane:
\begin{equation}
|z_{N}|\longrightarrow\infty
\label{eq:truncation-roots-escape}
\end{equation}
along an appropriate subsequence, or else disappear as the truncation order is changed.

A candidate pole that occurs only because:
\begin{equation}
F_{N}(z_{N})=0
\qquad\text{but}\qquad
F(z_{N})\neq0
\label{eq:false-doubler-condition}
\end{equation}
is a pole of the finite-order theory defined by $F_{N}$, not a pole of the complete theory defined by $F$. It must therefore be classified as a truncation artifact or false doubler.

A practical convergence test is to calculate the candidate roots for successive truncation orders:
\begin{equation}
z_{N},\ z_{N+1},\ z_{N+2},\ldots.
\end{equation}
A root that moves without convergence, leaves the controlled low-energy region, or escapes to infinity does not represent a finite particle pole of the exact nonlocal theory. The key test is always whether the pole occurs in the exact propagator $S_{\mathrm{ex}}(p)$.

\begin{theorem}[Mathematical and physical doubling test]
\label{thm:mathematical-physical-doubling-test}

Let $D_{\mathrm{ex}}(p)$ be the exact free inverse propagator of a translation-invariant fermion theory on a momentum manifold $\mathcal{M}$, and assume that $D_{\mathrm{ex}}(p)$ is continuously differentiable near its real finite-momentum zeros and possesses the analytic continuation required to define the Minkowski propagator.

An inequivalent momentum class $[p^{(j)}]$ represents a mathematical fermion species if and only if it satisfies the Dirac-regularity condition \eqref{eq:Dirac-regular-normal-form-test}. It represents a healthy physical fermion species if, in addition (1) it produces a nonremovable simple pole of the exact propagator; (2) the pole lies on a real mass shell with $m_{j}^{2}\geq0$; (3) its residue is nonzero and positive on the physical one-particle subspace; (4) the corresponding state is not removed by a momentum identification, gauge redundancy, constraint, or normalizability
condition; and (5) the pole is present in the exact untruncated free theory.

Every physical fermion doubler is associated with a mathematical Dirac-regular zero, but a mathematical zero need not represent a healthy physical fermion.
\end{theorem}

The proof of this theorem is trivial, we start by letting $[p^{(j)}]$ satisfy Eq.~\eqref{eq:Dirac-regular-normal-form-test}, then after nonsingular changes of basis the kinetic operator contains $m_{j}$ copies of a first-order Dirac operator. It thus contributes $m_{j}$ to the
mathematical species count \eqref{eq:mathematical-species-count}.

A physical asymptotic fermion must appear as a nonremovable pole of the propagator, for an isolated simple pole, the propagator has the form of \eqref{eq:physical-fermion-pole-form}. A real nonnegative mass shell, nonzero residue, and positive spectral weight are necessary for the pole to describe a stable positive-norm fermion state. Momentum-equivalent, constrained, gauge, and nonnormalizable modes do not define independent physical states. And finally, a pole absent from the exact untruncated operator is not a state of the exact theory. These conditions prove the stated classification we propose.

The alternatives may be summarized as:

\begin{widetext}
\begin{equation}
\begin{aligned}
\begin{split}
\begin{array}{c|c}
\text{Observed structure}
&
\text{Correct classification}
\\ \hline
\text{Additional inequivalent Dirac zero with a positive-residue pole}
&
\text{physical fermion doubler}
\\
\text{Dirac zero with negative residue}
&
\text{ghost-like mathematical mode}
\\
\text{Pole with }m^{2}<0
&
\text{tachyonic instability}
\\
\text{Complex pole}
&
\text{resonance or instability}
\\
\text{Zero cancelled in the full propagator}
&
\text{removable singularity}
\\
\text{Root present only in }F_{N}
&
\text{finite-truncation artifact}
\\
\text{Pole generated only by }\Sigma(p)
&
\text{dynamical excitation}
\\
\text{Momentum-equivalent zeros}
&
\text{one physical momentum mode}
\end{array}
\label{eq:doubling-classification-table}
\end{split}
\end{aligned}
\end{equation}
\end{widetext}

We will finally note an application to the continuum nonlocal Dirac operator. For the continuum nonlocal theory we have considered here, the exact free inverse propagator is:
\begin{equation}
D_{F}(p)
=
A(p)D_{0}(p),
\label{eq:nonlocal-operator-test-application}
\end{equation}
where:
\begin{equation}
D_{0}(p)=\slashed{p}-m
\label{eq:local-Dirac-operator-test-application}
\end{equation}
is the ordinary local Dirac operator and:
\begin{equation}
A(p)
=
F^{-1}\!\left(-\frac{p^{2}}{E_{M}^{2}}\right)
\label{eq:nonlocal-multiplier-test-application}
\end{equation}
is the nonlocal multiplier. The quantity $E_{M}>0$ is the nonlocality energy scale, and $F$ is assumed to be entire and nonvanishing at every finite argument. Since:
\begin{equation}
A(p)\neq0
\qquad
\forall\;\;\text{finite}\;\; p,
\label{eq:nonlocal-multiplier-nonzero-test}
\end{equation}
the multiplier is invertible. So:
\begin{equation}
D_{F}(p)u=0
\quad\Longleftrightarrow\quad
D_{0}(p)u=0
\label{eq:nonlocal-local-kernel-equivalence-test}
\end{equation}
for every spinor $u\in V$. Taking the determinant gives:
\begin{equation}
\det D_{F}(p)
=
A(p)^{\dim_{\mathbb{C}}V}
\det D_{0}(p),
\label{eq:nonlocal-determinant-factorization-test}
\end{equation}
where $\dim_{\mathbb{C}}V$ is the complex dimension of the spinor and internal vector space. Because the first factor in Eq.~\eqref{eq:nonlocal-determinant-factorization-test} never vanishes:
\begin{equation}
\det D_{F}(p)=0
\quad\Longleftrightarrow\quad
\det D_{0}(p)=0.
\label{eq:nonlocal-zero-set-equivalence-test}
\end{equation}
The nonlocal propagator is:
\begin{equation}
S_{F}(p)
=
iD_{F}(p)^{-1}
=
iD_{0}(p)^{-1}A(p)^{-1}.
\label{eq:correct-nonlocal-propagator-test}
\end{equation}
Using Eq.~\eqref{eq:nonlocal-multiplier-test-application}, this becomes:
\begin{equation}
S_{F}(p)
=
iD_{0}(p)^{-1}
F\!\left(-\frac{p^{2}}{E_{M}^{2}}\right).
\label{eq:nonlocal-propagator-form-factor-test}
\end{equation}
The entire form factor in Eq.~\eqref{eq:nonlocal-propagator-form-factor-test} is finite at every finite momentum and therefore cannot introduce an additional finite-momentum pole. Near the ordinary mass shell $p^{2}=m^{2}$:
\begin{equation}
S_{F}(p)
\sim
\frac{
 iF\!\left(-m^{2}/E_{M}^{2}\right)
 (\slashed{p}+m)
}{
 p^{2}-m^{2}+i\epsilon
}.
\label{eq:nonlocal-on-shell-residue-test}
\end{equation}
So the form factor changes the residue of the ordinary pole but does not create another pole.

If the physical reality condition:
\begin{equation}
F(x)\in\mathbb{R}
\qquad
\forall\;\;\text{real}\;\;x
\label{eq:form-factor-reality-condition}
\end{equation}
is imposed together with:
\begin{equation}
F(0)=1,
\qquad
F(x)\neq0
\quad
\forall\;\;\text{real}\;\;x,
\label{eq:real-zero-free-normalization}
\end{equation}
then continuity implies:
\begin{equation}
F(x)>0
\qquad
\forall\;\;\text{real}\;\;x.
\label{eq:form-factor-positive-real-axis}
\end{equation}
So a continuous real function beginning at $F(0)=1$ could become negative only by passing through zero, which is prohibited by Eq.~\eqref{eq:real-zero-free-normalization}. It then follows that the ordinary positive residue is not reversed by the nonlocal form factor.

The complete test therefore gives
\begin{equation}
N_{\mathrm{math}}
=
N_{\mathrm{phys}}
=
N_{\mathrm{int}}
\label{eq:nonlocal-final-species-equality}
\end{equation}
for the exact free continuum nonlocal Dirac theory. There are no additional inequivalent Dirac-regular zeros, no additional finite-momentum propagator poles, and thus no mathematical or physical fermion doubling.

\section{The No-Doubling Theorem}

We can now prove that the continuum nonlocal Dirac operator introduced above has exactly the same fermionic zero modes as the ordinary Dirac operator due to the fact that fermion doubling is the appearance of additional inequivalent zeros of the inverse propagator, equality of the two zero sets is sufficient to establish the result.

We recall from before:
\begin{equation}
\notag\mathcal{D}_F
=
A_F\mathcal{D}_0,
\qquad
A_F
=
F^{-1}\left(\frac{\Box_D}{E_M^2}\right),
\label{eq:factorized-nonlocal-Dirac}
\end{equation}
where again \(\mathcal{D}_0=i\gamma^\mu D_\mu-m\) is the ordinary covariant Dirac operator, \(\Box_D=D^\mu D_\mu\) is the covariant d'Alembertian, \(E_M>0\) is the nonlocal mass scale, and \(F\) is an entire function with no finite zeros. The operator \(A_F\) is therefore assumed to be invertible on the common domain of \(\mathcal{D}_0\) and \(\mathcal{D}_F\)~\cite{EvensMoffatKleppeWoodard1991,KleppeWoodard1992,ReedSimonI1980}.

\begin{theorem}
Let \(\mathcal{D}_0\) be a Dirac operator on a spinor space and let \(A_F\) be an invertible operator, if:
\begin{equation}
\mathcal{D}_F=A_F\mathcal{D}_0,
\label{eq:theorem-factorization}
\end{equation}
then:
\begin{equation}
\ker\mathcal{D}_F
=
\ker\mathcal{D}_0.
\label{eq:kernel-equality}
\end{equation}

Where \(\ker\mathcal{D}\) is the kernel or null space of the operator \(\mathcal{D}\), this is the set of spinor fields \(\psi\) satisfying \(\mathcal{D}\psi=0\).
\end{theorem}

For the proof we first suppose that:
\begin{equation}
\psi\in\ker\mathcal{D}_0,
\label{eq:psi-local-kernel}
\end{equation}
so by definition this means that:
\begin{equation}
\mathcal{D}_0\psi=0.
\label{eq:local-zero-mode}
\end{equation}
Using Eq.~\eqref{eq:theorem-factorization}, we then find:
\begin{equation}
\mathcal{D}_F\psi
=
A_F\mathcal{D}_0\psi
=
A_F0
=
0,
\label{eq:first-kernel-inclusion}
\end{equation}
and therefore every zero mode of the local Dirac operator is also a zero mode of the nonlocal operator:
\begin{equation}
\ker\mathcal{D}_0
\subseteq
\ker\mathcal{D}_F.
\label{eq:first-inclusion}
\end{equation}

Conversely, we suppose that:
\begin{equation}
\psi\in\ker\mathcal{D}_F,
\label{eq:psi-nonlocal-kernel}
\end{equation}
then:
\begin{equation}
A_F\mathcal{D}_0\psi=0,
\label{eq:nonlocal-zero-mode}
\end{equation}
and since \(A_F\) is invertible, we can act with \(A_F^{-1}\) on both sides:
\begin{equation}
A_F^{-1}A_F\mathcal{D}_0\psi
=
A_F^{-1}0.
\label{eq:apply-inverse}
\end{equation}
by using \(A_F^{-1}A_F=\mathbf{1}\), where \(\mathbf{1}\) is the identity operator, gives us:
\begin{equation}
\mathcal{D}_0\psi=0,
\label{eq:recover-local-equation}
\end{equation}
and hence:
\begin{equation}
\ker\mathcal{D}_F
\subseteq
\ker\mathcal{D}_0.
\label{eq:second-inclusion}
\end{equation}
By combining Eqs.~\eqref{eq:first-inclusion} and \eqref{eq:second-inclusion} proves Eq.~\eqref{eq:kernel-equality}.

This theorem is independent of whether \(A_F\) commutes with \(\mathcal{D}_0\). Only the ordering in Eq.~\eqref{eq:theorem-factorization} and the invertibility of \(A_F\) matter and are required. It thus applies both in the translation-invariant vacuum and, subject to the stated domain assumptions, in a nontrivial gauge background.

We now can specialize to flat spacetime where the momentum-space operator is:
\begin{equation}
\mathcal{D}_F(p)
=
F^{-1}\left(-\frac{p^2}{E_M^2}\right)
(\slashed p-m).
\label{eq:flat-nonlocal-Dirac-proof}
\end{equation}
where again \(p_\mu\) is the continuous four-momentum, \(p^2=\eta^{\mu\nu}p_\mu p_\nu\), \(\slashed p=\gamma^\mu p_\mu\), and \(m\) is the fermion mass. So since the scalar factor:
\begin{equation}
F^{-1}\left(-\frac{p^2}{E_M^2}\right)
\label{eq:scalar-entire-factor}
\end{equation}
is finite and nonzero for every finite \(p\), Eq.~\eqref{eq:flat-nonlocal-Dirac-proof} vanishes on a spinor \(u(p)\) \emph{iff}:
\begin{equation}
(\slashed p-m)u(p)=0,
\label{eq:ordinary-momentum-Dirac}
\end{equation}
so the nonlocal and local momentum-space operators have the same spinor solutions at every finite momentum.

The same conclusion will also follow from the determinant, for a four-component Dirac spinor:
\begin{equation}
\det(\slashed p-m)
=
(p^2-m^2)^2,
\label{eq:local-Dirac-determinant}
\end{equation}
the determinant vanishes when \(p^2=m^2\), which is the ordinary relativistic mass-shell condition. The square occurs because the four-dimensional Dirac matrix has two spin states for the particle sector and two for the antiparticle sector.

Since multiplication of a \(4\times4\) matrix by a scalar \(c\) multiplies its determinant by \(c^4\), Eq.~\eqref{eq:flat-nonlocal-Dirac-proof} gives:
\begin{equation}
\det\mathcal{D}_F(p)
=
F^{-4}\left(-\frac{p^2}{E_M^2}\right)
(p^2-m^2)^2,
\label{eq:nonlocal-Dirac-determinant}
\end{equation}
where the first factor in Eq.~\eqref{eq:nonlocal-Dirac-determinant} cannot vanish or diverge at finite momentum because \(F\) is entire and zero-free. So we see:
\begin{equation}
\det\mathcal{D}_F(p)=0
\quad\Longleftrightarrow\quad
p^2=m^2,
\label{eq:determinant-equivalence}
\end{equation}
and equation~\eqref{eq:determinant-equivalence} proves that the nonlocal operator possesses only the ordinary Dirac mass shell and no additional finite-momentum zeros.

The corresponding Feynman propagator is the inverse of the kinetic operator:
\begin{equation}
S_F(p)
=
iF\left(-\frac{p^2}{E_M^2}\right)
\frac{\slashed p+m}{p^2-m^2+i\epsilon},
\label{eq:nonlocal-fermion-propagator}
\end{equation}
where \(S_F(p)\) is the nonlocal fermion propagator and \(\epsilon>0\) specifies the usual Feynman boundary condition. The numerator contains the entire form factor and the matrix \(\slashed p+m\), while the denominator determines the propagator poles. Since the entire factor introduces no finite poles the only one-particle pole occurs at its usual point:
\begin{equation}
p^2=m^2,
\label{eq:single-physical-pole}
\end{equation}
the regulator may alter the residue and ultraviolet behavior of the propagator to keep it finite, but it does not alter the number of fermion poles.

For a massless fermion \(m=0\), the operator reduces to:
\begin{equation}
\mathcal{D}_F(p)
=
F^{-1}\left(-\frac{p^2}{E_M^2}\right)\slashed p,
\label{eq:massless-nonlocal-Dirac}
\end{equation}
its only finite zero is the ordinary continuum zero at:
\begin{equation}
p_\mu=0,
\label{eq:single-massless-zero}
\end{equation}
and there are no additional zeros at momenta analogous to the lattice values \(p_\mu=\pi/a\), because continuum momentum space is not periodic and the factor \(F(-p^2/E_M^2)\) is nonvanishing.

The key result is that a continuum Dirac theory deformed by multiplication with an invertible zero-free entire function of the covariant d'Alembertian has the same fermionic kernel, finite-momentum zero set, and one-particle pole set as the undeformed Dirac theory. So therefore the theory has no fermion-doubling problem.

\section{The 1991 Construction and the Recent Doubling Claim}

The relation between nonlocal regularization and fermion doubling was already discussed by Evens, Moffat, Kleppe, and Woodard in their 1991 construction of nonlocal gauge theories~\cite{EvensMoffatKleppeWoodard1991}. Their purpose was to obtain ultraviolet-finite, Poincar\'e-invariant, and perturbatively unitary gauge theories without changing either the particle content or the dimension of spacetime, a nontrivial feat. The construction introduced entire-function smearing operators together with nonlocal gauge transformations, higher-order interaction terms, and the functional measure required to preserve the corresponding quantum symmetry~\cite{Moffat1990,EvensMoffatKleppeWoodard1991,KleppeWoodard1992,Krasnikov1987}.

For a free fermion, the smearing operator used in that construction may be written schematically as:
\begin{equation}
\mathcal{E}
=
\exp\left(\frac{\mathcal{K}}{2\Lambda^2}\right),
\label{eq:1991-smearing-operator}
\end{equation}
where \(\mathcal{E}\) is the nonlocal smearing operator, \(\mathcal{K}\) is the local quadratic kinetic operator, and \(\Lambda>0\) is the nonlocal regularization scale. The precise sign in the exponential depends on the Minkowski and Euclidean conventions but after Wick rotation the form factor is chosen to suppress large Euclidean momentum.

In the fermion sector, the local kinetic operator is given by:
\begin{equation}
\mathcal{K}_{\psi}
=
i\slashed{\partial}-m,
\label{eq:1991-local-fermion-operator}
\end{equation}
where \(\slashed{\partial}=\gamma^\mu\partial_\mu\), \(\gamma^\mu\) are the Dirac gamma matrices, and \(m\) is the fermion mass. The smeared fermion field is then defined by an analytic function of the kinetic operator:
\begin{equation}
\widehat{\psi}
=
\mathcal{E}^{-1}\psi,
\label{eq:1991-smeared-fermion}
\end{equation}
here \(\psi\) is the unsmeared Dirac field and \(\widehat{\psi}\) is its nonlocal counterpart. Because the exponential function has no finite zeros, \(\mathcal{E}\) is invertible on the domain on which the functional calculus is defined.

The 1991 paper argued that the massless nonlocal theory could preserve chiral symmetry without producing the species doubling familiar from lattice regularization~\cite{EvensMoffatKleppeWoodard1991,NielsenNinomiyaNoGo1981,NielsenNinomiyaI1981,NielsenNinomiyaII1981}. The authors correctly observed that the Nielsen–-Ninomiya theorem assumes locality and therefore does not directly apply to their nonlocal construction. This identifies the assumption that removes the theory from the scope of the lattice no-go theorem, but it does not by itself prove that no additional zeros occur in the exact fermion kinetic operator.

The theorem established in the preceding section gives the necessary spectral proof of the no doubling theorem. If the nonlocal fermion operator has the factorized form:
\begin{equation}
\mathcal{D}_{\rm NL}
=
\mathcal{E}^{-1}\mathcal{D}_0,
\label{eq:1991-factorized-Dirac}
\end{equation}
where \(\mathcal{D}_0=i\slashed{\partial}-m\) is the ordinary Dirac operator and \(\mathcal{E}^{-1}\) is invertible, then:
\begin{equation}
\ker\mathcal{D}_{\rm NL}
=
\ker\mathcal{D}_0,
\label{eq:1991-kernel-equality}
\end{equation}
this equation means that the set of spinor fields annihilated by the nonlocal operator is exactly the same as the set annihilated by the local Dirac operator. The nonlocal smearing therefore cannot create additional fermionic zero modes.

In a translation-invariant vacuum, the operator becomes:
\begin{equation}
\mathcal{D}_{\rm NL}(p)
=
\mathcal{E}^{-1}(p)(\slashed p-m),
\label{eq:1991-momentum-Dirac}
\end{equation}
where \(p_\mu\) is the continuous four-momentum and \(\mathcal{E}(p)\) is the momentum-space entire-function multiplier. Since:
\begin{equation}
\mathcal{E}(p)\neq 0
\label{eq:1991-nonzero-smearing}
\end{equation}
for every finite momentum, the determinant satisfies:
\begin{equation}
\det\mathcal{D}_{\rm NL}(p)
=
\mathcal{E}^{-4}(p)(p^2-m^2)^2,
\label{eq:1991-determinant}
\end{equation}
here the fourth power appears because the Dirac operator acts on four-component spinors. Equation~\eqref{eq:1991-determinant} vanishes only on the ordinary mass shell \(p^2=m^2\). So the 1991 construction contains no additional finite-momentum Dirac zeros and no Nielsen–-Ninomiya fermion doublers.

A recent comment by Cline states, in a footnote attributed to a private communication from Woodard, that the earlier claim in the 1991 paper of avoiding fermion doubling was mistaken, as noted before this was a misinterpretation on Clines part of what Woodard stated, but the comment still remains public on arXiv. In this comment no explicit modified Dirac operator or additional zero of the fermion inverse propagator was presented, so the assertion therefore cannot be assessed independently of the exact operator being considered, and is false~\cite{Cline2025,ClineComment,MoffatThompsonReply2025}.

For the continuum entire-function construction analyzed here a contrary result would require a finite momentum \(p^{(j)}\) satisfying:
\begin{equation}
\det\mathcal{D}_{\rm NL}\left(p^{(j)}\right)=0,
\qquad
\left(p^{(j)}\right)^2\neq m^2,
\label{eq:hypothetical-extra-zero}
\end{equation}
where \(p^{(j)}\) is a hypothetical additional fermionic branch, with the index \(j\) distinguishing it from the ordinary Dirac solution. Equation~\eqref{eq:1991-determinant} excludes such a momentum whenever the entire form factor is finite, zero-free, and invertible.

The later analysis carried out by Clayton, Demopoulos, and Moffat clarified an important point that is logically separate from fermion doubling, they showed that the original nonlocal construction had suggested that the axial anomaly might be absent. The later work showed instead that nonlocal QED possesses the ordinary axial anomaly once the transformation of the functional measure is included, the regulated triangle amplitudes and the measure contribution reproduce the standard anomalous Ward identity in the local limit~\cite{Adler1969,BellJackiw1969,Fujikawa1979,ClaytonDemopoulosMoffat1994}.

This result does not overturn the no-doubling statement as an axial anomaly is the quantum failure of a classically defined axial current to remain conserved, whereas fermion doubling is the appearance of additional inequivalent zeros of the fermion inverse propagator. The existence of the anomaly thus neither proves nor requires additional fermion species.

Indeed, the propagators used in the 1993 analysis preserve the ordinary physical pole and the smeared propagator is obtained by multiplying the local propagator by a nonvanishing exponential entire function, and hence has the same finite pole set. The accompanying shadow propagator has its apparent local pole canceled by its numerator, the shadow field therefore has no propagating pole and is excluded from the asymptotic state space and it cannot be interpreted as a fermion doubler.

The measure analysis reinforces this point, the quantum measure is required to remain entire in the momentum invariants precisely so that no additional degrees of freedom are excited. When the authors attempt to gauge the full anomalous chiral symmetry, they find that preserving it would require either new pole structure or additional fields. This is an obstruction to quantizing the fully gauged chiral theory, not evidence that the original nonlocal fermion spectrum is doubled.

The statement in that work that fermion doubling would remove the anomaly in the massless limit is conditional as it describes the anomaly cancellation that additional mirror species could produce as it does not establish that such species are present. On the contrary actually, the nonzero anomaly obtained in the calculation shows that no hidden set of doublers automatically cancels the axial contribution.

We should thus say that within the assumptions of the 1991 nonlocal construction the original conclusion is correct, that abandoning strict locality removes the theory from the Nielsen–-Ninomiya hypotheses, while invertibility of the entire smearing operator ensures directly that the fermion spectrum is not multiplied.

\section{On Finite Truncations and Interactions}

The no-doubling theorem applies to the complete nonlocal operator defined by the full entire function and this distinction is very important because the infinite derivative expansion and any finite truncation of that expansion generally define different theories. This is the same kind of thing that happens with MP3 players, since audio is stored as continuous numbers representing a wave~\cite{BrandenburgBosi1997}. When a file is truncated a mathematical function is instantly cut off at a non-zero value, so a smooth wave function \(f(x)\) is suddenly replaced by a step function dropping instantly to \(0\). In calculus this creates a sharp, non-differentiable corner (a discontinuity) and to a speaker this instantaneous drop to zero requires infinite acceleration, causing a violent physical "pop" or "chirp." To convert digital data into sound, an MP3 player uses algorithms based on the Fourier Transform, which breaks down audio into pure sine waves~\cite{ISOIEC11172-3,BrandenburgStoll1994,Shlien1994,Pan1995,Noll1997,Brandenburg1999,PainterSpanias2000,BosiGoldberg2003,Herre2019}. Smooth curves require few frequencies and a sharp mathematical cliff (the truncation) requires an infinite sum of high-frequency sine waves to represent it mathematically, this is known as the Gibbs phenomenon~\cite{Wilbraham1848,Gibbs1899,Gibbs1898,Bocher1906,HewittHewitt1979,GottliebShu1997,Carslaw1925}. The audio result is that the player tries to decode these sudden, infinite high frequencies and because the math forces an impossible shape, it produces high-pitched, metallic digital noise, this is the "chirp." MP3 data is mathematically divided into discrete blocks (the frames) of 1,152 audio samples. Now if a file is truncated mid-frame, the player is missing the final variables needed to solve the decompression equation~\cite{PrincenBradley1986,PrincenJohnsonBradley1987}. So the result for the audio is that the decoder is forced to perform math on incomplete arrays. It fills the missing slots with "null" data or random bits, translating the broken math into audio garbage.

We recall that the entire form factor can be written as:
\begin{equation}
\notag F(z)
=
\sum_{n=0}^{\infty}c_n z^n,
\label{eq:full-entire-expansion}
\end{equation}
where as before \(z\in\mathbb{C}\) is a complex variable and \(c_n\) are the Taylor coefficients of \(F\). In the nonlocal fermion theory the argument is the dimensionless operator:
\begin{equation}
z
=
\frac{\Box_D}{E_M^2},
\label{eq:dimensionless-operator-argument}
\end{equation}
where \(\Box_D=D^\mu D_\mu\) is the gauge-covariant d'Alembertian and \(E_M>0\) is the nonlocal mass scale. The complete operator is thus:
\begin{equation}
 F\left(\frac{\Box_D}{E_M^2}\right)
=
\sum_{n=0}^{\infty}
c_n
\left(\frac{\Box_D}{E_M^2}\right)^n,
\label{eq:complete-nonlocal-operator}
\end{equation}
where equation~\eqref{eq:complete-nonlocal-operator} is an analytic operator. Its infinitely many derivative terms are not independent propagating fields and must not be analyzed separately~\cite{BarnabyKamran2008,BuoninfanteLambiaseMazumdar2019,ModestoRachwal2017}.

For example, the exponential choice of $F(z)=e^{-z}$ is entire and satisfies:
\begin{equation}
\notag F(z)\neq 0
\qquad
\forall\;\;\text{finite}\;z\in\mathbb{C},
\label{eq:exponential-zero-free}
\end{equation}
and its reciprocal \(F^{-1}(z)=e^{z}\) is therefore also entire. So trivially multiplication of the Dirac operator by either \(F\) or \(F^{-1}\) cannot introduce any new finite-momentum zeros.

A finite truncation is instead defined by:
\begin{equation}
F_N(z)
=
\sum_{n=0}^{N}c_n z^n,
\label{eq:finite-truncation}
\end{equation}
where \(N\) is a fixed nonnegative integer. We note that unlike \(F\), the function \(F_N\) is a polynomial. A nonconstant polynomial generally has complex roots \(z_j\) satisfying:
\begin{equation}
F_N(z_j)=0,
\label{eq:truncation-roots}
\end{equation}
if we were to replace the exact form factor by \(F_N\), these roots may produce additional zeros or poles at momenta satisfying:
\begin{equation}
-\frac{p^2}{E_M^2}=z_j,
\label{eq:spurious-momentum-roots}
\end{equation}
where \(p_\mu\) is the continuum four-momentum and \(p^2=\eta^{\mu\nu}p_\mu p_\nu\). These roots are artifacts of the finite-order approximation and they are not present in the full zero-free entire function and are not Nielsen–-Ninomiya doublers because they do not arise from a periodic Brillouin zone or from multiple lattice Dirac cones.

The exact no-doubling statement is then:
\begin{equation}
F(z)\neq 0
\quad\Longrightarrow\quad
Z(\mathcal{D}_F)=Z(\mathcal{D}_0),
\label{eq:exact-zero-set-statement}
\end{equation}
where \(Z(\mathcal{D})\) is the finite-momentum zero set of the operator \(\mathcal{D}\). No equivalent statement is guaranteed for an arbitrary polynomial approximation \(F_N\).

These zeros are not genuine doublers as they only appear due to finite truncations, these are called false doublers or pseudo-doublers. This is not a new theorem, but we are writing it in a somewhat new context.

In the theorem below we will make a temporary notational simplification, earlier the exact nonlocal Dirac operator was written as:
\[
D_F(p)=F^{-1}\left(-\frac{p^2}{E_M^2}\right)D_0(p),
\]
and since $(F)$ is entire and zero-free, its reciprocal $(F^{-1})$ is also entire and zero-free. For the purposes of the following theorem only, we relabel the entire kinetic multiplier $(F^{-1})$ as $(F)$. So, throughout the theorem and its corollary:
\[
F_{\mathrm{theorem}}(z)
:=
F_{\mathrm{earlier}}^{-1}(z).
\]
This change is purely notational and does not affect the kernel, zero-set, or pole arguments. The polynomial $(F_N)$ appearing below is therefore the Taylor truncation of the kinetic multiplier $(F_{\mathrm{earlier}}^{-1})$, not the Taylor truncation of the earlier function $(F_{\mathrm{earlier}})$.

\begin{theorem}[False-doubling theorem for finite derivative truncations]
Let \(E_M>0\) be a nonlocality scale, and let the admissibility conditions be satisfied~\eqref{eq:admissible-form-factor}. If we have an entire function~\eqref{eq:entire-map} satisfying a globally convergent power-series expansion~\eqref{eq:entire-series}
possesses a globally convergent power-series expansion. And new we let the exact translation-invariant nonlocal Dirac operator be:
\begin{equation}
D_F(p)
=
F\!\left(-\frac{p^2}{E_M^2}\right)D_0(p),
\label{eq:false-doubling-exact-operator}
\end{equation}
where $D_0(p)=\slashed p-m$ is the ordinary continuum Dirac operator, \(p\in\mathbb R^{1,d-1}\), \(m\geq 0\), and \(\slashed p=\gamma^\mu p_\mu\).

We then let:
\begin{equation}
F_N(z)
=
\sum_{n=0}^{N}
\frac{F^{(n)}(0)}{n!}z^n
\label{eq:false-doubling-truncation}
\end{equation}
be the Taylor polynomial of \(F\) truncated at a finite order \(N\), and define the corresponding finite-derivative operator:
\begin{equation}
D_N(p)
=
F_N\!\left(-\frac{p^2}{E_M^2}\right)D_0(p).
\label{eq:false-doubling-truncated-operator}
\end{equation}
Then the following statements will hold; firstly the exact nonlocal operator has exactly the same finite-momentum zero set and kernel as the local Dirac operator:
\begin{equation}
\ker D_F(p)=\ker D_0(p),
\qquad
Z(D_F)=Z(D_0),
\label{eq:false-doubling-kernel}
\end{equation}
where we have:
\[
Z(D):=\left\{p\in\mathbb R^{1,d-1}:\det D(p)=0\right\}.
\]
The next thing is that the truncated operator has the zero set:
\begin{equation}
Z(D_N)
=
Z(D_0)
\cup
\left\{
p\in\mathbb R^{1,d-1}:
F_N\!\left(-\frac{p^2}{E_M^2}\right)=0
\right\}.
\label{eq:false-doubling-truncated-zero-set}
\end{equation}
The third point is that every momentum \(p_\star\) satisfying:
\begin{equation}
F_N\!\left(-\frac{p_\star^2}{E_M^2}\right)=0,
\qquad
\det D_0(p_\star)\neq 0,
\label{eq:false-doubling-spurious-root}
\end{equation}
is a zero of the finite-order truncated operator but not of the exact nonlocal operator. Any apparent fermionic excitation inferred from such a zero is therefore a truncation-induced false mode and not a degree of freedom of the complete nonlocal theory. And finially that such truncation-induced zeros are not Nielsen–-Ninomiya fermion doublers since they arise from the algebraic roots of the polynomial \(F_N\), rather than from distinct topologically required zeros of a periodic lattice Dirac operator on a compact Brillouin zone.
\end{theorem}

The proof is trivial as \(F(z)\neq 0\) at every finite complex argument, the scalar factor is nonzero for every finite momentum \(p\). It is therefore invertible as a finite-dimensional scalar multiplier at each momentum. So we have that $D_F(p)u=0$ and is is equivalent to
\[
F\!\left(-\frac{p^2}{E_M^2}\right)D_0(p)u=0,
\]
which, by multiplication with the inverse scalar factor, is equivalent to $D_0(p)u=0$. So we have that:
\[
\ker D_F(p)=\ker D_0(p)
\]
for every finite \(p\), proving the equality of the kernels. Taking determinants in \(d_s\)-dimensional spinor space gives us:
\begin{equation}
\det D_F(p)
=
F\!\left(-\frac{p^2}{E_M^2}\right)^{d_s}
\det D_0(p).
\label{eq:false-doubling-exact-determinant}
\end{equation}
Now since the entire-function factor is nonzero, we have:
\[
\det D_F(p)=0
\quad\Longleftrightarrow\quad
\det D_0(p)=0.
\]
Therefore we see $Z(D_F)=Z(D_0)$.

Now we see for the truncated operator:
\begin{equation}
\det D_N(p)
=
F_N\!\left(-\frac{p^2}{E_M^2}\right)^{d_s}
\det D_0(p).
\label{eq:false-doubling-truncated-determinant}
\end{equation}
So then \(\det D_N(p)=0\) whenever either $\det D_0(p)=0,$ or $F_N\!\left(-\frac{p^2}{E_M^2}\right)=0$. So this proves Eq.~\eqref{eq:false-doubling-truncated-zero-set}.

Now we let \(p_\star\) satisfy Eq.~\eqref{eq:false-doubling-spurious-root}, then we see that $\det D_N(p_\star)=0$, whereas
\[
\det D_F(p_\star)
=
F\!\left(-\frac{p_\star^2}{E_M^2}\right)^{d_s}
\det D_0(p_\star)
\neq 0.
\]
So \(p_\star\) belongs to the zero set of the truncated operator but not to the zero set of the exact operator when we use the full Taylor expansion. The corresponding apparent excitation is thus produced just by replacing the complete entire function with a finite polynomial.

Finally we can say that the Nielsen–-Ninomiya doubling concerns the topology of a periodic lattice Dirac operator defined over a compact Brillouin torus. But the additional zeros described above instead depend on the roots of the chosen polynomial \(F_N\), vary with the truncation order \(N\), and disappear when the exact zero-free entire function is restored. So they are just truncation artifacts rather than genuine lattice fermion doublers.

\begin{corollary}[The Absence of false doubling in the complete theory]
Under the assumptions of the theorem, the complete nonlocal propagator
\begin{equation}
\notag S_F(p)
=
iD_F(p)^{-1}
=
iD_0(p)^{-1}
F\!\left(-\frac{p^2}{E_M^2}\right)^{-1}
\label{eq:false-doubling-exact-propagator}
\end{equation}
has no additional finite-momentum poles beyond those already present in the local Dirac propagator corresponding to physical fermions.

But the formally resummed truncated propagator:
\begin{equation}
S_N(p)
=
iD_0(p)^{-1}
F_N\!\left(-\frac{p^2}{E_M^2}\right)^{-1}
\label{eq:false-doubling-truncated-propagator}
\end{equation}
may possess additional poles at the roots of \(F_N\). These poles belong to the finite-order higher-derivative theory defined by \(D_N\), not to the complete entire-function theory defined by \(D_F\).
\end{corollary}

Now after this corollary we go back to the original convention in which the nonlocal kinetic operator is written as $(F^{-1}D_0)$.

One final remark, the polynomial \(F_N\) may be used consistently as an order-by-order low-energy approximation when:
\[
\left|\frac{p^2}{E_M^2}\right|\ll 1.
\]
But we should note its roots must not be interpreted as exact particle states when they occur at momenta for which the omitted higher-order terms are no longer negligible. Treating the finite Taylor polynomial as an exact kinetic function changes the theory and can manufacture additional real, complex, tachyonic, or ghost-like poles that are absent from the original zero-free entire-function operator~\cite{BarnabyKamran2008,BiswasGerwickKoivistoMazumdar2012,BuoninfanteLambiaseMazumdar2019,ThompsonOstrogradsky2026}.

Now after discussing the false doubling we now can consider interactions, we know that the exact one-particle content of an interacting fermion theory is determined by the zeros of the full one-particle-irreducible two-point function and in momentum space, we write the exact inverse propagator as:
\begin{equation}
\Gamma_F^{(2)}(p)
=
F^{-1}\left(-\frac{p^2}{E_M^2}\right)
(\slashed p-m)
-
\Sigma_F(p),
\label{eq:interacting-inverse-propagator}
\end{equation}
where \(\Gamma_F^{(2)}(p)\) is the exact inverse fermion propagator and \(\Sigma_F(p)\) is the nonlocal fermion self-energy generated by interactions. The quantity \(\slashed p=\gamma^\mu p_\mu\) is the Dirac contraction of momentum with the gamma matrices, and \(m\) is the bare fermion mass.

The physical poles are determined by:
\begin{equation}
\det\Gamma_F^{(2)}(p)=0,
\label{eq:interacting-pole-condition}
\end{equation}
now interactions may shift the physical mass and wave-function residue, as they do in ordinary quantum field theory. They may also generate bound-state or resonance poles in special models but the point is that such dynamically generated poles are not fermion doubling. Fermion doubling refers specifically to an unavoidable multiplicity of regulator-induced kinetic zeros already present in the free theory.

The result established in the previous section is a kinematical statement:
\begin{equation}
\Sigma_F(p)=0
\label{eq:free-limit-pole-result}
\end{equation}
$\det\Gamma_F^{(2)}(p)=0$ only when $p^2=m^2$. So the entire-function deformation itself does not generate additional fermion species and any extra pole in an interacting theory would have to arise from the dynamics contained in \(\Sigma_F(p)\), but is not from the nonlocal regulator or from a Nielsen–-Ninomiya doubling mechanism.

So if the continuum nonlocal theory is subsequently discretized on a lattice, the lattice Dirac operator must be analyzed separately. A naive lattice replacement of the derivative can reintroduce the usual periodic zeros even when an entire-function factor is also present. That would be a property of the chosen lattice discretization and not of the underlying continuum nonlocal quantum field theory.

\section{Final Comment on the Recent Doubling Claim}

In a recent comment by Cline\footnote{As noted before this was a misinterpretation of what Woodard stated.}, he states that the earlier claim that nonlocal regularization avoids fermion doubling was mistaken, with the supporting reference given as a private communication from Woodard~\cite{Cline2025,ClineComment,MoffatThompsonReply2025}. Since no explicit doubled spectrum is presented the assertion must be tested against the mathematical criterion established above.

For a free fermion doubling requires additional inequivalent zeros of the inverse propagator. In the continuum nonlocal theory considered here, the inverse propagator is:
\begin{equation}
\notag \mathcal{D}_F(p)
=
F^{-1}\left(-\frac{p^2}{E_M^2}\right)
(\slashed p-m),
\label{eq:claim-test-operator}
\end{equation}
where \(p_\mu\) is the continuous four-momentum, \(p^2=\eta^{\mu\nu}p_\mu p_\nu\), \(m\) is the fermion mass, \(\slashed p=\gamma^\mu p_\mu\), \(E_M>0\) is the nonlocal mass scale, and \(F\) is entire and nonvanishing at finite argument.

To establish a doubling problem, he or anyone else would have to exhibit at least one additional finite momentum \(p^{(j)}\), inequivalent to the ordinary Dirac mass shell, such that:
\begin{equation}
\det\mathcal{D}_F\!\left(p^{(j)}\right)=0,
\label{eq:required-extra-zero}
\end{equation}
where here the label \(j\) distinguishes a hypothetical additional fermion zero. Such a zero would represent an extra low-energy fermionic excitation only if the operator linearized around \(p^{(j)}\) had the form of an independent Dirac operator.

However, from the determinant identity:
\begin{equation}
\det\mathcal{D}_F(p)
=
F^{-4}\left(-\frac{p^2}{E_M^2}\right)
(p^2-m^2)^2,
\label{eq:claim-determinant}
\end{equation}
and the assumption:
\begin{equation}
\notag F(z)\neq 0
\qquad
\forall\;\;\text{finite }z\in\mathbb{C},
\label{eq:claim-zero-free}
\end{equation}
it follows that:
\begin{equation}
\det\mathcal{D}_F(p)=0
\quad\Longleftrightarrow\quad
p^2=m^2.
\label{eq:claim-refutation}
\end{equation}
And thus no additional finite-momentum zero exists in the full continuum theory.

The claim could only apply to a different construction like for example maybe a finite polynomial truncation that may possess some artificial roots, and a naive lattice discretization may produce the usual Brillouin-zone doublers. But neither case is equivalent to the exact continuum operator in Eq.~\eqref{eq:claim-test-operator}.

It is also important to separate the doubling question from gauge invariance, a nonlocal gauge theory must be completed so that the relevant Ward or Slavnov--Taylor identities are preserved, and failure to construct the required gauge-completed interactions may lead to gauge-symmetry violations, but it does not by itself imply fermion doubling. Doubling is again a statement about additional zeros of the fermion kinetic operator, while gauge consistency is a statement about the symmetry relations among propagators, vertices, and amplitudes.

We can therefore confidently conclude that the recent assertion does not establish a fermion-doubling problem for zero-free entire-function nonlocal quantum field theory. A valid counterexample to the theorem proved above would require either a finite zero of the form factor, a failure of operator invertibility on the stated domain, or an explicitly demonstrated additional zero of the exact inverse propagator. None follows merely from the nonlocality of the continuum theory.

\section{Concluding Remarks}

In this paper we have explored the fermion-doubling question directly from the spectral structure of the continuum nonlocal Dirac operator, here the relevant criterion is not just the presence of nonlocality or infinitely many derivatives but it is whether the exact inverse propagator develops additional inequivalent zeros corresponding to new low-energy fermionic excitations.

For the class of nonlocal quantum field theories we study the ordinary Dirac operator is multiplied by an entire-function operator that is nonvanishing and invertible on the relevant domain, with this we proved the multiplication such an operator preserves the kernel of the Dirac operator and in the translation-invariant vacuum, this means that the nonlocal and local kinetic operators have exactly the same finite-momentum zero set, thus maintaining the fermion sector of local quantum theory. The corresponding nonlocal propagator keeps only the ordinary Dirac particle pole and contains no regulator-induced fermion species.

This result is conceptually distinct from the Nielsen–-Ninomiya theorem as that theorem concerns local lattice fermions defined over a compact and periodic Brillouin zone~\cite{NielsenNinomiyaNoGo1981,NielsenNinomiyaI1981,NielsenNinomiyaII1981,Friedan1982,ChandrasekharanWiese2004,Kaplan2009}. The topological constraint forces the zeros of a chiral lattice Dirac operator to occur in compensating sets. We know that nonlocal quantum field theory is not formulated on a momentum torus and is not local in the sense assumed by the theorem, so it then therefore falls outside the theorem's hypotheses, and more importantly the absence of doubling does not follow merely from this failure of the lattice assumptions but from the direct equality of the local and nonlocal kernels proved in this paper.

We also have distinguished the complete nonlocal theory we study from the idea of taking finite derivative truncations~\cite{BarnabyKamran2008,BuoninfanteLambiaseMazumdar2019,ThompsonOstrogradsky2026}. The full infinite series maintains a zero-free entire function, whereas a finite polynomial approximation may possess artificial roots and may introduce spurious higher-derivative modes, but these are not real poles in the propagator. And these roots belong to the truncated theory and do not establish a doubling problem in the exact nonlocal theory that we study. Similarly a later discretization of the continuum theory may introduce ordinary lattice doublers if a naive finite-difference Dirac operator is used but that would be a property of the lattice mapping and not that of the underlying continuum formulation.

The recent statement that nonlocal regularization fails to avoid fermion doubling was not presented with an explicit additional zero of the exact nonlocal Dirac operator. Within the assumptions stated here such a zero cannot occur unless the form factor ceases to be invertible or develops a finite zero. The burden of proof for any contrary claim is thus to identify a specific exact kinetic operator and exhibit the additional fermionic branches in its spectrum.

The conclusion of this paper is precise and limited, we showed that in a continuum Dirac theory deformed by a zero-free entire function of a covariant differential operator does not suffer from fermion doubling. Nonlocality changes the ultraviolet behavior and position-space support of the fermion field, but it does not multiply the fermion species. We also gave tests to determine if the doubling is real or false and how to test if it is mathematical or physical.

A final thought we want to leave you with is that for decades fermion doubling has been a pathology of lattice theories, sometimes it is good and sometimes it is bad. Fermion doubling acts as both a protective shield for standard physics and an obstacle for engineering next-generation quantum devices. This fermion doubling can be good because it ensures physical stability, respects time-reversal symmetry, and provides distinct valley states for processing quantum information. However, it can also be bad because in systems like graphene the paired valleys perfectly cancel out unique single-valley topological effects and allow defects to cause performance-degrading intervalley scattering. But for models of quantum gravity and fundamental theories it was thought that this effect was inherently bad, but in this paper we have given tests and theorems that can truly help us understand the nature of this pathology. The thought we want to leave you with is that even though physicists thought we understood this theorem we have not truly understood how to apply it, so we should in future focus on understanding at a foundational level what the pathologies are telling us.

\section*{Acknowledgments}

We thank James Cline for helpful discussions and writing the comment that sparked this paper and Richard Woodard for the original comment made to Cline (even though it was a misinterpretation by Cline). We thank Martin Green for helpful comments and discussions on the paper. We would also like to thank Hilary Carteret for discussions. Research at the Perimeter Institute for Theoretical Physics is supported by the Government of Canada through Industry Canada and by the Province of Ontario through the Ministry of Research and Innovation (MRI).
\renewcommand{\notesname}{Endnotes}
\theendnotes

\end{document}